\documentclass[a4paper,12pt]{article}
\pdfoutput=1 

\usepackage{jheppub}
\usepackage{graphicx}
\usepackage{axodraw}
\usepackage{epsfig}
\usepackage{amsmath, amsthm, amssymb}
\usepackage{amsfonts}
\usepackage{subfig}
\usepackage{color}
\usepackage{xspace}
\usepackage[dvipsnames]{xcolor}
\usepackage{paralist}
\usepackage{multirow}
\usepackage{paralist}
\usepackage{slashed}

\newcommand{\beq}{\begin{equation}}
\newcommand{\eeq}{\end{equation}}
\newcommand{\bea}{\begin{eqnarray}}
\newcommand{\eea}{\end{eqnarray}}
\newcommand{\bfig}{\begin{figure}}
\newcommand{\efig}{\end{figure}}
\newcommand{\bc}{\begin{center}}
\newcommand{\ec}{\end{center}}

\newcommand{\lsim}{\lesssim}

\newcommand{\gsim}{\gtrsim}

\def\s2tw{{\rm sin ^2 \theta_{W}}}

\def\sq2{\sqrt{2}}
\newcommand{\be}{\begin{equation}}
\newcommand{\ee}{\end{equation}}

\preprint{IIPDM-2018-10}

\title{Neutrino Masses in a Two Higgs Doublet Model with a U(1) Gauge Symmetry}

\author[a]{Daniel A. Camargo,}
\author[b]{Alex G. Dias,}
\author[a,c]{T\'{e}ssio B. de Melo,}
\author[a]{Farinaldo S.\ Queiroz}

\affiliation[a]{International Institute of Physics, Universidade Federal do Rio Grande do Norte, Campus Universitario, Lagoa Nova, Natal-RN 59078-970, Brazil}

\affiliation[b]{Universidade Federal do ABC, Centro de Ci\^{e}ncias Naturais e Humanas, Santo Andr\'{e},SP, Brazil}

\affiliation[c]{Departamento de F\'{\i}sica, Universidade Federal da Para\'\i ba, Caixa Postal 5008, 58051-970, Jo\~ao Pessoa, PB, Brazil}

\emailAdd{farinaldo.queiroz@iip.ufrn.br}

\abstract{General Two Higgs Doublet Models (2HDM) are popular Standard Model extensions but feature flavor changing interactions and lack neutrino masses. We discuss a 2HDM where neutrino masses are generated via type I seesaw and propose an extension where neutrino masses are generated via a type II seesaw mechanism with flavor changing interactions being absent via the presence of a U(1) gauge symmetry. After considering a variety of bounds such as those rising from collider and electroweak precision we show that our proposal stands as a UV complete 2HDM with a dark photon where neutrino masses and flavor changing interactions are addressed. A possible dark matter realization is also discussed.}


\begin{document} 
\maketitle
\flushbottom

\clearpage
\section{Introduction}
\label{sec:1}

The Standard Model (SM) is the most accurate description of nature to the electroweak and strong interactions \cite{Glashow:1961tr,Weinberg:1967tq,Tanabashi:2018oca}. The discovery of a 125 GeV spin-0 state at CERN was the last piece of the puzzle in the SM \cite{Aad:2012tfae,Chatrchyan:2012xdj} and established the existence of, as far as we know, an elementary scalar particle in nature. However, elementary scalar particles are common figures in many beyond the SM adventures, and among those Two-Higgs-Doublet stand out \cite{Lee:1973iz}. The $\rho$ parameter, $\rho=m_W^2/(m_Z^2 \cos ^2 \theta _W)$ is a powerful probe to those models featuring multiple scalar particles because they may contribute to the gauge boson masses \cite{Branco:2011iw} and, therefore, alter the SM prediction. The current value from global fits point to  $\rho=1.00039\pm 0.00019$ \cite{Tanabashi:2018oca}. The gauge boson masses arise from the kinetic terms of the scalars, thus the $\rho$ parameter can be parametrized at tree level as,

\begin{equation}
\rho = \frac{{\displaystyle \sum_{i=1}^n} \left[
I_i \left( I_i+1 \right) - \frac{1}{4}\, Y_i^2 \right] v_i}
{{\displaystyle \sum_{i=1}^n}\, \frac{1}{2}\, Y_i^2 v_i}, 
\label{tessio1}
\end{equation}where $I_i$, $Y_i$ and $v_i$ are the isospins and hypercharges and vacuum expectation values of the scalars. From eq. (\ref{tessio1}) we can see that scalar doublets ($I=1/2$) with $Y=\pm 1$ and scalar singlets ($I=0$) with $Y=0$ do not contribute to the $\rho$ parameter, and for this reason are desired extensions of the SM. \\

Two-Higgs-Doublet Models (2HDM) have indeed proven to be interesting models featuring a rich phenomenology concerning collider physics \cite{Davidson:2010sf,Nomura:2017wxf,Camargo:2018klg}, axion models \cite{Dasgupta:2013cwa,Alves:2016bib}, baryogenesis \cite{Turok:1990zg,Cline:1995dg,Clarke:2015hta}, flavor physics \cite{Botella:2011ne,Ko:2012sv,Davidson:2016utf,Gaitan:2017tka,Martinez:2018ynq}, among others \cite{Xu:2017vpq,Chen:2018uim}. Several versions of 2HDM have been proposed in the literature trying to improve the original proposal in some theoretical aspects, via the inclusion of dark matter \cite{LopezHonorez:2006gr,Gustafsson:2007pc,Dolle:2009fn,Goudelis:2013uca,Honorez:2010re,LopezHonorez:2010tb,Arhrib:2013ela,Bonilla:2014xba,Queiroz:2015utg,Arcadi:2017wqi} and neutrino masses \cite{Antusch:2001vn,Atwood:2005bf,Chao:2012pt,Liu:2016mpf,Cheung:2017lpv,Bertuzzo:2018ftf}. A proposal to explain neutrino masses in the context of 2HDM has already been put forth with no connection to gauge symmetries and absence of flavor changing neutral interactions (FCNI) \cite{Ma:1998dx,Ma:2000cc,Ma:2002nn,Grimus:2009mm}. A common feature in these studies is the presence of an ad-hoc discrete symmetry where one of the scalar doublets is odd under, which is added to avoid FCNI. It would be theoretically elegant if all these problems 
that general 2HDMs face could be solved in connection to gauge symmetries. Nevertheless, some proposals to extend the 2HDM via the presence of gauge symmetries have been put forth \cite{Ko:2012hd,Huang:2015wts,Arhrib:2018sbz}. Some were triggered by anomalies in flavor and collider physics \cite{Heeck:2014qea,Crivellin:2015mga,DelleRose:2017xil} and others devoted to explain neutrino masses via type I seesaw mechanism and absence of flavor changing interactions \cite{Ko:2013zsa,Ko:2014uka,Ko:2015fxa,Campos:2017dgc}. \\

In this work, we propose a new model, still within the scope of 2HDM but different from previous studies. We explain the absence of FCNI via an abelian gauge symmetry $U(1) _X$ and neutrino masses via type II seesaw mechanism \cite{Mohapatra:1979ia,Schechter:1980gr,Mohapatra:1980yp}. The presence of a new abelian gauge symmetry gives rise to gauge anomalies which, in order to be canceled out, impose restrictions over the SM fermion charges under the new symmetry. Moreover, this abelian group induces the presence of a $Z^\prime$ gauge boson. The scalar doublets develop vacuum expectation value (VEV) at the electroweak scale and the VEV of the scalar triplet cannot be large due to bounds stemming from the $\rho$ parameter, hence the $Z^\prime$ is necessarily light, with masses below the weak scale. This fact has important experimental implications which will be  investigated. We will also briefly discuss how one could potentially accommodate a dark matter candidate in our model.\\ 

Our work is structured as follows: In section \ref{typeImodel}, we introduce the 2HDM-U(1) model and revisit how neutrino masses could be easily generated via a type I seesaw mechanism; In section \ref{typeIImodel} we show how to implement a type II seesaw mechanism and obtain the mass spectrum of the model; In section \ref{pheno} we discuss some phenomenological constraints; In section \ref{sec:concl} we draw our conclusions. At the end we left three sections in the appendix where details of the anomaly cancellation and spontaneous symmetry breaking mechanism are shown.

\section{Type I seesaw in the 2HDM-U(1)}
\label{typeImodel}

As aforementioned, general 2HDM suffer from severe bounds rising from flavor physics. The problem of FCNI at tree level can be elegantly handled by the introduction of an extra abelian gauge symmetry. This new gauge symmetry is certainly more theoretically appealing than the usually ad-hoc $ Z _2 $ discrete symmetry which must be explicitly broken in order to not generate domain walls \cite{Zeldovich:1974uw}. We will now revisit how the addition of an abelian gauge symmetry benefits 2HDM and generates neutrino masses via type I seesaw mechanism \cite{Minkowski:1977sc,Das:2012ii,Alonso:2012ji,Deppisch:2015qwa}. In this model, we have two scalar doublets $ \Phi _i \sim ( 1 , 2 , 1 , Q _{X _i} ) $ with the same hypercharge $ Y = 1 $ where,
\begin{equation}
\Phi _i = \begin{pmatrix} \phi ^+ \\ \phi ^0 \end{pmatrix} \text{\ \ \ , \ \ \ \ } \phi ^0 = \frac{\rho _i + v _i + i \eta _i}{\sqrt{2}} \text{\ \ \ , \ \ \ \ } i = 1, 2 ,
\end{equation}with $Q_{X_1}\neq Q_{X_2}$. The fact that $Q_{X_1}\neq Q_{X_2}$ leads to the scalar potential, 

\begin{equation}
\begin{split}
\label{potential_doblets_u1_x}
V_d  & = m _1 ^2 \Phi _1 ^\dagger \Phi _1 + m _2 ^2 \Phi _2 ^\dagger \Phi _2 + \frac{\lambda _1}{2} ( \Phi _1 ^\dagger \Phi _1 ) ^2 + \frac{\lambda _2}{2} ( \Phi _2 ^\dagger \Phi _2 ) ^2 \\
& + \lambda _3 ( \Phi _1 ^\dagger \Phi _1 ) ( \Phi _2 ^\dagger \Phi _2 ) + \lambda _4 ( \Phi _1 ^\dagger \Phi _2 ) ( \Phi _2 ^\dagger \Phi _1 ) .
\end{split}
\end{equation}

Since the scalar doublets have different charges under $U(1) _X$ only one of them will couple to SM fermions, and we arbitrarily choose $ \Phi _2 $. In this way, we get the Yukawa Lagrangian, 

\begin{equation}
\label{yukawa_2hdm_typeI}
- \mathcal{L} _{ Y _{\text{2HDM-I}} } = y _2 ^d \bar{Q} _L \Phi _2 d _R + y _2 ^u \bar{Q} _L \tilde{\Phi} _2 u _R + y _2 ^e \bar{L} _L \Phi _2 e _R + h.c. 
\end{equation}

Thus far our model is nearly identical to the usual type I 2HDM model proposed previously in the literature. The key difference lies in the introduction of a gauge symmetry which naturally explain the origin of the $Z_2$ symmetry. Since we also want to accommodate neutrino masses via type I seesaw mechanism we need to add three right-handed neutrinos as follows,

\begin{equation}
\label{yukawa_2hdm_typeI_right_neutrinos}
- \mathcal{L} _{ Y _{N _R} } = y _2 ^D \bar{L} _L \tilde{\Phi} _2 N _R + y ^M \overline{N _R ^c} \Phi _s N _R + h.c. 
\end{equation}
where we now included a singlet scalar $\Phi_s$, charged under $U(1) _X$, to build a majorana mass term, which features a scalar potential,

\begin{equation}
\begin{split}
\label{potential_doblets_singlet_u1_x}
V _s & = m _s ^2 \Phi _s ^\dagger \Phi _s + \frac{\lambda _s}{2} ( \Phi _s ^\dagger \Phi _s ) ^2 + \mu ( \Phi _1 ^\dagger \Phi _2 \Phi _s + h. c. ) \\
& + \lambda _{s1} ( \Phi _1 ^\dagger \Phi _1 ) ( \Phi _s ^\dagger \Phi _s ) + \lambda _{s2} ( \Phi _2 ^\dagger \Phi _2 ) ( \Phi _s ^\dagger \Phi _s ) .
\end{split}
\end{equation}

In the Appendix \ref{sec:appvcb} we show the vacuum stability bounds for the potential $V=V_d+V_s$ given by eqs. (\ref{potential_doblets_u1_x}) and (\ref{potential_doblets_singlet_u1_x}). The perturbative unitarity bounds, as well as the vacuum stability bounds, for a similar potential  containing two Higgs doublets plus a scalar singlet are presented in Ref. \cite{Arhrib:2018qmw}. 

With these Lagrangians we can simultaneously explain neutrino masses and the absence of FCNI. However, there is still one point that needs to be addressed which is the presence of gauge anomalies arising from the introduction of a new gauge symmetry. Since the SM fermions can be charged under the new gauge group and we have introduced new chiral fermions (right-handed neutrinos), the anomaly cancellation procedure becomes non-trivial. Generally the $\Phi_2$ charge under $U(1) _X$ is different from zero, and therefore the SM fermions should also be charged under $U(1) _X$. The anomaly cancellation procedure is described in Appendix \ref{sec:app0}. We highlight that we can in principle keep the model anomaly free without the addition of right-handed neutrinos. However, without them we would not be able to explain neutrino masses the way we wished for. Anyways, with their presence we can derive the anomaly cancellation requirements that preserve our Lagrangians as follows,

\begin{equation}
\begin{split}
\label{expres_cargas_u_d}
q = \frac{1}{2} ( u + d & ) \text{\ \ \ , \ \ \ \ } l = - \frac{3}{2} ( u + d ) \text{\ \ \ , \ \ \ \ } e = - ( 2 u + d ) \text{\ \ \ , \ \ \ \ } n = - ( u + 2 d ) , \\
Q _{X _1} & = \frac{1}{2} ( 5 u + 7 d ) \text{\ \ \ , \ \ \ \ } Q _{X _2} = \frac{1}{2} ( u - d ) \text{\ \ \ , \ \ \ \ } q _X = 2 u + 4 d ,
\end{split}
\end{equation}where $u$ and $d$ are the $U(1) _X$ charges of the up and down quarks respectively, $ q $ ($ l $) the charge of the quark (lepton) doublet, $ e $ ($n$) the charge of the right-handed charged leptons (neutrinos), and lastly $ Q _{X _i} $ ($ q _X $) the $U(1) _X$ charge of the scalar doublet (singlet).

The scalar singlet $\Phi_s$ is responsible for generating a majorana mass term for the right-handed neutrinos and breaking the $U(1) _X$ gauge symmetry that yields a massive $Z^\prime$ gauge boson. A natural question that rises to this conclusion is: is there a way to explain neutrino masses without adding a singlet scalar and right-handed neutrinos within the scope of 2HDM? Motivated by this question we will describe hereafter how one could accomplish that.

\section{Type II seesaw in the 2HDM-U(1)}
\label{typeIImodel}

A popular mechanism to explain the active neutrino masses without the presence of right-handed neutrinos  is the so called type II seesaw mechanism \cite{Magg:1980ut,Lazarides:1980nt}. In order to implement this mechanism within the scope of 2HDM the fermion charges under the gauge $U(1) _X$ symmetry need to be tied to one another to cancel out the triangle anomalies.
We have seen above that one of the anomaly cancellation conditions is $n=-(u+2d)$, which comes from the $U(1)^3$ triangle anomaly as shown in the Appendix \ref{sec:app0}, where $n$ is the right-handed neutrino charge under $U(1) _X$. Therefore, if there are no right-handed neutrinos we must set $ u = - 2d $ to be free from gauge anomalies. Compared to the type I seesaw scenario, instead of having two independent charges ($u$ and $d$), we now have only one, say $d$. That implies into,

\begin{equation}
\begin{split}
\label{expres_cargas_d}
q & = - \frac{d}{2} \text{\ \ , \ \ \ } l = \frac{3d}{2} \text{\ \ , \ \ \ } e = 3 d , \\
u & = - 2 d \text{\ \ \ \ , \ \ \ \ \ } Q _{X _2} = - \frac{3d}{2} .
\end{split}
\end{equation}

The charge of the first doublet is free, as long as $ Q _{X _1} \neq Q _{X _2} $, in order to recover the Yukawa Lagrangian (\ref{yukawa_2hdm_typeI}) and keep the model free from FCNI. As shown in the Table \ref{cargas_u1_2hdm_tipoI}, there is essentially only two different possibilities. One where the SM fermions are neutral and the other where they are charged under $U(1) _X$. If a particular nonzero value is chosen for $d$, any other multiple of this value would produce a physically equivalent model, because a change in $d$ can be balanced by a rescaling on the gauge coupling constant $g _X$, so that the $U(1) _X$ interaction remains the same. In particular, taking $d=-2/3$ we notice that the charges of the SM fermions under $U(1) _X$ are similar to the SM weak hypercharge. In this way, it is clear that a type II seesaw realization in the 2HDM-U(1) gives rise either to a fermiophobic or a sequential $Z^\prime$ boson.\\

\begin{table}[!t]
\centering
{
\bf Charges in Type II seesaw 2HDMs free from FCNI}
\begin{tabular}{ccccccccc}
\hline 
Fields & $u_R$ & $d_R$ & $Q_L$ & $L_L$ & $e_R$ & $\Delta$ & $\Phi _2$ & $\Phi_1$ \\ \hline 
Charges & $-2d$ & $d$ & $-d/2$ & $3d/2$ & $3d$ & $-3d$ & $-3d/2$ & $\neq Q _{X _2} $  \\
$U(1)_{N}$ & $0$ & $0$ & $0$ & $0$ & $0$ & $0$ & $0$  &  $\neq Q _{X _2} $\\
$U(1)_{Y^\prime} $ & $4/3$ & $-2/3$ & $1/3$ & $-1$ & $-2$ & $2$ & $1$ & $\neq Q _{X _2} $ \\
\end{tabular}
\caption{The table shows anomaly free Type I 2HDM where neutrino masses are generated via a type II seesaw mechanism. The first row shows the generic charges as functions of the $d _R$ quark charge, $d$. Two particular cases are shown for $d = 0$ and $d = - 2/3$, which correspond to sequential $Z'$ and dark photon models, respectively. Notice that to prevent FCNI the scalar doublets have different charges under the $U(1) _X$ gauge symmetry. }
\label{cargas_u1_2hdm_tipoI}
\end{table}

The implementation of type II seesaw mechanism requires an $SU (2) _L$ scalar triplet $\Delta \sim ( 1 , 3 , 2 , q _{X _t} ) $, where the quantum numbers refers to the transformation properties under the symmetry group $ SU(3) _c \times SU(2) _L \times U(1) _Y \times U(1) _X $. The field $ \Delta $ can be parameterized as,
\begin{equation}
\Delta = \begin{pmatrix} \Delta ^+ / \sqrt{2} & \Delta ^{++} \\ \Delta ^0 & - \Delta ^+ / \sqrt{2} \end{pmatrix} ,
\end{equation}
with,
\begin{equation}
\Delta ^0 = \frac{\rho _t + v _t + i \eta _t}{\sqrt{2}} .
\end{equation}
The $ SU(2) _L $ symmetry allows the introduction of an interaction between $ \Delta $ and the leptons via,
\begin{equation}
\label{yukawa_lepton_triplet}
- \mathcal{L} _{Y _t} = y  _L \overline{L _L ^c} i \sigma ^2 \Delta L _L + h.c.
\end{equation}which requires $\Delta$ to have hypercharge $Y _t = 2$ and lepton number $ L _t = 2 $, automatically forbidding interactions to quarks. The inclusion of eq. (\ref{yukawa_lepton_triplet}) implies,
\begin{equation}
\label{yukawa_constraint_u_1_triplet_charge}
2l + q _{X _t} = 0 ,
\end{equation}and using eq. (\ref{expres_cargas_d}) we get,
\begin{equation}
q _{X _t} = - 3 d,
\end{equation}
explaining the $\Delta$ charge shown in Table \ref{cargas_u1_2hdm_tipoI}. \\

Since $\Delta$ carries lepton number, when the neutral scalar $\Delta^0$ develops a VEV, $v_t$, lepton number is violated, and from eq. (\ref{yukawa_lepton_triplet}) we can easily see that it generates a majorana mass term for the neutrinos with,

\begin{equation}
m _\nu = \sqrt{2} y  _L v _t .
\label{Eqneutrinomass}
\end{equation}

Thus, $ v _t $ has to be very small in order to accommodate neutrino masses in the sub-eV range. In summary, with the presence of a $U(1) _X$ gauge symmetry, we can explain the absence of FCNI and accommodate neutrino masses via a type II seesaw without extra fermions, which is the main idea of this work. However, we need also to study the phenomenological implications of such proposal before concluding whether we have a feasible theoretical model. We start studying the mass spectrum of the model.

\subsection{Mass Spectrum - Scalars}

Our goal in this section is to study the spontaneous symmetry breaking pattern to find the physical scalars, gauge bosons, and neutrino masses. The SM charged lepton masses are the same as in the SM. That said, we begin our reasoning with the scalar sector.\\

The scalar sector is described by the Lagrangian,
\begin{equation}
\label{lag_scalar_sector}
\mathcal{L} _{\text{scalar}} = ( D _\mu \Phi _i ) ^\dagger ( D ^\mu \Phi _i ) + \text{Tr} [ ( D _\mu \Delta ) ^\dagger ( D ^\mu \Delta ) ] - V ( \Phi _1, \Phi _2 , \Delta ) ,
\end{equation}where the covariant derivatives of the scalar doublets and the triplet read,
\begin{equation}
\label{covariant_derivative_1}
D _\mu \Phi _i = \partial _\mu \Phi _i + i g \tau ^a W _\mu ^a + i g ' \frac{Y}{2} \hat{B} _\mu \Phi _i + i g _X \frac{Q _{X _i}}{2} \hat{X} _\mu \Phi _i ,
\end{equation}
\begin{equation}
\label{covariant_derivative_2}
D _\mu \Delta = \partial _\mu \Delta + i g [ \tau ^a W _\mu ^a , \Delta ] + i g ' \frac{Y _t}{2} \hat{B} _\mu \Delta + i g _X \frac{q _{X _t}}{2} \hat{X} _\mu \Delta ,
\end{equation}where $ \tau ^a $ are the generators of the $ SU(2) _L $ group. The scalar potential in eq. (\ref{lag_scalar_sector}), invariant under all gauge symmetries is given by
\begin{equation}
\begin{split}
V ( \Phi _1 ,  \Phi _2 , \Delta ) & = m _1 ^2 \Phi _1 ^\dagger \Phi _1 + m _2 ^2 \Phi _2 ^\dagger \Phi _2 + m _t ^2 \text{Tr} ( \Delta ^\dagger \Delta ) + \lambda _1 ( \Phi _1 ^\dagger \Phi _1 ) ^2 + \lambda _2 ( \Phi _2 ^\dagger \Phi _2 ) ^2 \\
& + \lambda _3 ( \Phi _1 ^\dagger \Phi _1 ) ( \Phi _2 ^\dagger \Phi _2 ) + \lambda _4 ( \Phi _1 ^\dagger \Phi _2 ) ( \Phi _2 ^\dagger \Phi _1 ) + \lambda _{t1} ( \Phi _1 ^\dagger \Phi _1 ) \text{Tr} ( \Delta ^\dagger \Delta ) \\
& + \lambda _{t2} ( \Phi _2 ^\dagger \Phi _2 ) \text{Tr} ( \Delta ^\dagger \Delta ) + \lambda _{tt1} \Phi _1 ^\dagger \Delta \Delta ^\dagger \Phi _1 + \lambda _{tt2} \Phi _2 ^\dagger \Delta \Delta ^\dagger \Phi _2 \\
& + \lambda _t [ \text{Tr} ( \Delta ^\dagger \Delta ) ] ^2 + \lambda _{tt} \text{Tr} ( \Delta ^\dagger \Delta ) ^2 + \mu _{t2} ( \Phi _2 ^T i \sigma ^2 \Delta ^\dagger \Phi _2 + h.c. ) .
\label{scalarpotential}
\end{split}
\end{equation}\\

The necessary conditions for having vacuum stability with this potential are given in Appendix \ref{sec:appvcb}. Due the presence of the scalar triplet, the unitarity bounds are much more involved than the potential with the scalar singlet in section \ref{typeImodel}. We postpone an complete analysis of the unitarity and vacuum stability bounds to another work.

Observe in the potential above that the terms $ \Phi _1 ^T i \sigma ^2 \Delta ^\dagger \Phi _2 $ and $\Phi _1 ^T i \sigma ^2 \Delta ^\dagger \Phi _1 $ are forbidden by the $U(1)_X$ symmetry, as we require $ Q _{X1} \neq Q _{X2} $. There is only one non-hermitian term $\Phi _2 ^T i \sigma ^2 \Delta ^\dagger \Phi _2 $, which breaks lepton number in two units. Such lepton number violation is a common feature in seesaw type II models. It is important to note that neutrino masses are generated when $\Delta^0$ develops a vacuum expectation value as shown in eq.(\ref{Eqneutrinomass}) and that would be related to lepton number violation since the scalar triplet carries lepton number. However, notice that the non-hermitian term in eq.(\ref{scalarpotential}) already explicitly violates lepton number, thus lepton number had been violated even before $\Delta^0$ develops a non-trivial vacuum expect value. We checked that without this non-hermitian term in the scalar potential the pseudoscalar from the scalar triplet field would remain massless, i.e. a majoron field \cite{Queiroz:2014yna}.\\

Anyway, substituting the VEVs,
\begin{equation}
\langle \phi ^0 _i \rangle = \frac{v _i}{\sqrt{2}}\,,\,\,\,\,\,\,\, \langle \Delta ^0 \rangle = \frac{v _t}{\sqrt{2}} ,
\end{equation}
in order to break spontaneously the gauge symmetries, we have the following constraint equations for a minimal point of the potential, \\ \\
\begin{equation}
m _1 ^2 + \frac{1}{2} \left[ 2 \lambda _1 v _1 ^2 + ( \lambda _3 + \lambda _4 ) v _2 ^2 + ( \lambda _{t1} + \lambda _{tt1} ) v _t ^2 \right] = 0 ,
\end{equation}
\begin{equation}
m _2 ^2 + \frac{1}{2} \left[ ( \lambda _3 + \lambda _4 ) v _1 ^2 + 2 \lambda _2 v _2 ^2 + ( \lambda _{t2} + \lambda _{tt2} ) v _t ^2 - 2 \sqrt{2} \mu _{t2} v _t \right] = 0 ,
\end{equation}
\begin{equation}
\label{vinc}
\frac{v _t}{2} \left[2m _t ^2 + ( \lambda _{t1} + \lambda _{tt1} ) v _1 ^2 + ( \lambda _{t2} + \lambda _{tt2} ) v _2 ^2 + 2 ( \lambda _t + \lambda _{tt} ) v _t ^2 \right]-  \frac{\mu _{t2} v _2 ^2}{\sqrt{2}} = 0 .
\end{equation}

In the Standard Model, the symmetry is spontaneously broken when the mass term flips sign. In the case of a type II seesaw, the mass term of the scalar triplet in the scalar potential does not need to flip sign to break the symmetry. We emphasize that the gauge symmetry is broken when the scalar doublets acquire a non-zero vacuum expectation value. See \cite{Arhrib:2011uy} for a detailed discussion about the type II seesaw vacuum.  Moreover, we will see later on that the mass term of the scalar triplet should be positive in order to generate a pseudoscalar with positive mass. 

Assuming that $2m_t^2$ is the dominant term between the brackets in the constraint equation (\ref{vinc}), we have a seesaw relation, 
\begin{equation}
\label{seesaw_II_relation}
v _t \simeq \frac{\mu _{t2} v _2 ^2}{\sqrt{2} m _t ^2} ,
\end{equation}
which leads to a naturally dwindled $ v _t $ for $|m _t^2|\gg |\mu _{t2} v _2|$. In this way, a small $v_t$ can be understood as a simply consequence of having the coefficient of  the bilinear term in $\Delta$ to be comparatively large with respect to the other energy scales of the scalar potential. Note that from eq. (\ref{seesaw_II_relation}) we conclude that $m_t^2$ and $\mu_{t2}$ should have the same sign. \\

In the scalar sector, $ \Phi _i $ and $ \Delta $ render the existence of seven physical fields: 3 CP-even scalars, $ h $, $ H $ and $ H _t $; one CP-odd, $ A $; two singly charged $ H ^+ $, $ H _t ^+ $ and one doubly charged $ H ^{++} $. The other scalar degrees of freedom are absorbed as longitudinal components by the gauge bosons, $ W ^{\pm} $, $ Z $ and $ Z ' $, making them massive. \\

In the basis $ ( \rho _1 , \rho _2 , \rho _t ) $ the neutral scalars mix according to the following mass matrix, 
\begin{equation}
\label{cp_even_mass_matrix}
M ^2 _{\text{CPeven}} = \begin{pmatrix}  2 \lambda _1 v _1 ^2 & ( \lambda _3 + \lambda _4 ) v _1 v _2 & ( \lambda _{t1} + \lambda _{tt1} ) v _1 v _t \\
 ( \lambda _3 + \lambda _4 ) v _1 v _2 & 2 \lambda _2 v _2 ^2 & ( \lambda _{t2} + \lambda _{tt2} ) v _2 v _t - \sqrt{2} \mu _{t2} v _2 \\
 ( \lambda _{t1} + \lambda _{tt1} ) v _1 v _t & ( \lambda _{t2} + \lambda _{tt2} ) v _2 v _t - \sqrt{2} \mu _{t2} v _2 & 2 ( \lambda _t + \lambda _{tt} ) v _t ^2 + \frac{\mu _{t2} v _2^2}{\sqrt{2} v _t}
\end{pmatrix} . 
\end{equation}\\

From the diagonalization procedure of this mass matrix we will get three physical scalars, $ h $, $ H $ and $ H _t $. We can parametrize this diagonalization in terms of three mixing angles $ \alpha $, $ \alpha _1 $ and $ \alpha _2 $, 
\begin{equation}
\begin{split}
\label{rot_to_phy_basis_cp_even}
\begin{pmatrix} h \\ H \\ H _t \end{pmatrix} = & \begin{pmatrix} c _\alpha & s _\alpha & 0 \\ - s _\alpha & c _\alpha & 0 \\ 0 & 0 & 1 \end{pmatrix} 
\begin{pmatrix} c _{\alpha _1} & 0 & s _{\alpha _1} \\ 0 & 1 & 0 \\ - s _{\alpha _1} & 0 & c _{\alpha _1} \end{pmatrix} \begin{pmatrix} 1 & 0 & 0 \\ 0 & c _{\alpha _2} & s _{\alpha _2} \\ 0 & - s _{\alpha _2} & c _{\alpha _2} \end{pmatrix} \begin{pmatrix} \rho _1 \\ \rho _2 \\ \rho _t \end{pmatrix} ,
\end{split}
\end{equation}where $ s _{\alpha,\alpha_1,\alpha_2} $ and $ c _{\alpha,\alpha_1,\alpha_2} $ are sine and cosine functions. We choose $ h $ to denote the $ 125 $ GeV SM-like Higgs found in the LHC \cite{Chatrchyan:2012xdj,Aad:2012tfae}. The angles are determined by the parameters of the potential and the scalar VEVs. Fully analytic expressions for the masses and eigenvectors are complicated but we can obtain approximate results. As shown in the Appendix \ref{sec:app1}, in the limit $v _t \ll v _i$, the masses of the CP-even scalars are approximately,
\begin{equation}
\label{mass_neutral_approximate_1}
m _{h} ^2 = \lambda _1 v _1 ^2 + \lambda _2 v _2 ^2 - \sqrt{ ( \lambda _1 v _1 ^2 - \lambda _2 v _2 ^2 )^2 + ( \lambda _3 + \lambda _4 ) ^2 v _1 ^2 v _2 ^2 } - 2 \sqrt{2} \sin ^2 \alpha \text{\ } \mu _{t2} v _t 
\end{equation}
\begin{equation}
\label{mass_neutral_approximate_2}
m _{H} ^2 = \lambda _1 v _1 ^2 + \lambda _2 v _2 ^2 + \sqrt{ ( \lambda _1 v _1 ^2 - \lambda _2 v _2 ^2 )^2 + ( \lambda _3 + \lambda _4 ) ^2 v _1 ^2 v _2 ^2 } - 2 \sqrt{2} \cos ^2 \alpha \text{\ } \mu _{t2} v _t
\end{equation} 
\begin{equation}
\label{masses_neutral_approximate_3}
m _{H _t} ^2 = \frac{\mu _{t2} v _2 ^2}{\sqrt{2} v _t} .
\end{equation} \\

\begin{figure}[!t]
\begin{center}
    \includegraphics[width=\columnwidth]{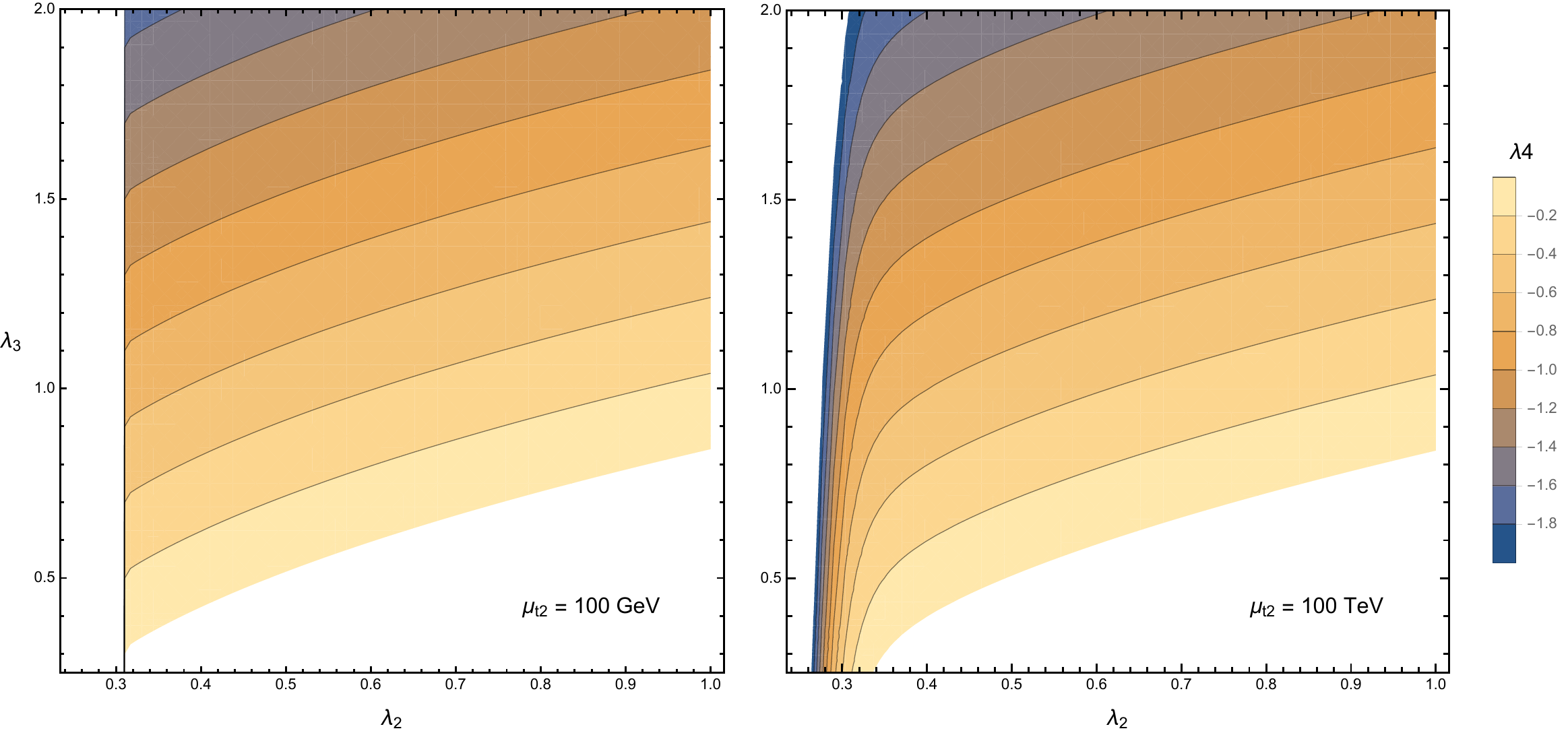}
\caption{Region of parameter space that leads to a $125$ GeV SM-like Higgs boson for $ v _2 = 200 $ GeV, $ v _t = 1$ MeV and $ \lambda _1 = 0.6 $. In the left panel $ \mu _{t2} = 100 $ GeV and in the right panel, $ \mu _{t2} = 100 $ TeV. }
\label{SM_like_Higgs_parameter_space}
\end{center}
\end{figure}

This limit  $v _t \ll v _i$ that will be assumed throughout this work yields a higgs boson, $h$, with the correct mass is shown in Fig. \ref{SM_like_Higgs_parameter_space}.  It is straightforward to see that we can easily find a higgs with the correct mass for couplings of order one and $\mu_{t2}$ either at the weak of multi-TeV scale.
\\

As for the pseudoscalars, in the basis $ ( \eta _1 , \eta _2 , \eta _t ) $ the mass matrix is given by,
\begin{equation}
M ^2 _{\text{CPodd}} = \sqrt{2} \mu _{t2} \begin{pmatrix} 0 & 0 & 0 \\
 0 & 2 v _t & - v _2 \\
 0 & - v _2 & \frac{v _2 ^2 }{2 v _t} 
\end{pmatrix} .
\end{equation}
Note that $ \eta _1 $ is decoupled and massless. Thus it can be immediately recognized as a Goldstone boson, $G_1$. After the diagonalization procedure we found another Goldstone boson, $G_2$. These two massless pseudoscalars represent the degrees of freedom needed to generate the $Z$ and $Z^\prime$ masses. In the diagonalization procedure we find the rotation matrix,

\begin{equation}
\begin{split}
    \begin{pmatrix} G _1 \\ G _2 \\ A \end{pmatrix} = & \begin{pmatrix} 1 & 0 & 0 \\ 0 & c _{\beta '} & s _{\beta '} \\ 0 & - s _{\beta '} & c _{\beta '} \end{pmatrix} \begin{pmatrix} \eta _1 \\ \eta _2 \\ \eta _t \end{pmatrix},
\end{split}
\end{equation}where,
\begin{equation}
\tan \beta ' = \frac{2 v _t}{v _2} ,
\end{equation}which gives rise to two massless fields as aforementioned and a massive pseudoscalar, $A$, with mass, 
\begin{equation}
m _A ^2 = \frac{\mu _{t2} ( v _2 ^2 + 4 v _t^2 ) }{\sqrt{2} v _t} .
\label{eqpseudoA}
\end{equation}\\

Observe that $v _t$ and $ \mu _{t2} $ must have the same sign in order to have $ m _A ^2 > 0 $. We had concluded previously from eq.(\ref{seesaw_II_relation}) that $\mu_{t2}$ and $m_t^2$ should have the same sign to keep $v_t$ positive definite, thus from eq.(\ref{eqpseudoA}) $\mu_{t2}$ must be positive to generate a positive squared mass for the pseudoscalar $A$. Hence, both $\mu_{t2}$ and $m_t^2$ are strictly positive.
\\

It is important to stress that even with the introduction of a new gauge symmetry, the pseudoscalar $A$, which is a common figure in 2HDM, remains in the spectrum. Under the assumption that $v_t$ is smaller than $\mu_{t2}$ the pseudoscalar can have a mass sufficiently large to evade existing bounds, as we shall discuss further. \\

The charged scalars mass matrix in the basis $ ( \phi _1 ^+ , \phi _2 ^+ , \Delta ^+ ) $ is,
\begin{equation}
\label{charged_scalars_mass_matrix}
\small{ 
M _{\text{Charged}} ^2 = \frac{1}{2} \begin{pmatrix} - \lambda _4 v _2 ^2 - \lambda _{tt1} v _t ^2 & \lambda _4 v _1 v _2 & \lambda _{tt1} v _1 v _t / \sqrt{2} \\
 \lambda _4 v _1 v _2 & - \lambda _4 v _1 ^2 - \lambda _{tt2} v _t ^2 + 2 \sqrt{2} \mu _{t2} v _t & \frac{1}{2} ( \sqrt{2} \lambda _{tt2} v _t - 4 \mu _{t2} ) v _2 \\
 v _1 v _t \lambda _{tt1} / \sqrt{2} & \frac{1}{2} ( \sqrt{2} \lambda _{tt2} v _t - 4 \mu _{t2} ) v _2 & \frac{ \sqrt{2} \mu _{t2} v _2 ^2 }{v _t} - \frac{1}{2} ( \lambda _{tt1} v _1 ^2 + \lambda _{tt2} v _2 ^2 )
\end{pmatrix}.
}
\end{equation}

The physical fields are given by performing the following rotation,
\begin{equation}
\begin{split}
\label{rot_to_phy_basis_charged_scalars}
\begin{pmatrix} G ^+ \\ H ^+ \\ H _t ^+ \end{pmatrix} = & \begin{pmatrix} c _\beta & s _\beta & 0 \\ - s _\beta & c _\beta & 0 \\ 0 & 0 & 1 \end{pmatrix} 
\begin{pmatrix} c _{\beta _1} & 0 & s _{\beta _1} \\ 0 & 1 & 0 \\ - s _{\beta _1} & 0 & c _{\beta _1} \end{pmatrix} \begin{pmatrix} 1 & 0 & 0 \\ 0 & c _{\beta _2} & s _{\beta _2} \\ 0 & - s _{\beta _2} & c _{\beta _2} \end{pmatrix} \begin{pmatrix} \phi _1 ^+ \\ \phi _2 ^+ \\ \Delta ^+ \end{pmatrix} .
\end{split}
\end{equation}\\

The Goldstone boson $G ^+ $ is absorbed by $ W ^+ $, and the physical states $ H ^+ $ and $ H _t ^+ $ have masses,
\begin{equation}
\label{masses_charged_exact_1}
m _{H ^+} ^2 = \frac{1}{8} ( A - \sqrt{A ^2 - B} ),
\end{equation}
\begin{equation}
\label{masses_charged_exact_2}
m _{H _t ^+} ^2 = \frac{1}{8} ( A + \sqrt{A ^2 - B} ) ,
\end{equation}
where,
\begin{equation*}
A = - 2 \lambda _4 \left( v _1 ^2 + v _2 ^2 \right) - \lambda _{tt1} \left( v _1 ^2 + 2 v _t ^2 \right) - \lambda _{tt2} \left( v _2 ^2 + 2 v _t ^2 \right) + 2 \sqrt{2} \frac{\mu _{t2}}{v _t} \left( v _2 ^2 + 2 v _t ^2 \right),
\end{equation*}
\begin{equation*}
B = 8 ( v _1 ^2 + v _2 ^2 + 2 v _t ^2 ) \left[ \lambda _4 \left( \lambda _{tt1} v _1 ^2 + \lambda _{tt2} v _2 ^2 \right) + \lambda _{tt1} \lambda _{tt2} v _t ^2 - 2 \sqrt{2} \frac{\mu _{t2}}{v _t} \left( \lambda _4 v _2 ^2 + \lambda _{tt1} v _t ^2 \right) \right] .
\end{equation*}\\

The doubly charged scalar $ \Delta ^{++} $ does not mix any other field. This mass eigenstate, which we will denote henceforth by $ H^{++} $, has a mass given by,
\begin{equation}
\label{eqdoubly}
m ^2 _{H ^{\pm \pm}} = \frac{ \mu _{t2} v _2 ^2 }{\sqrt{2} v _t} - \frac{1}{2} ( \lambda _{tt1} v _1 ^2 + \lambda _{tt2} v _2 ^2 + 2 \lambda _{tt} v _t ^2 ) .
\end{equation} \\

In summary, the scalar mass spectrum is largely controlled by the relative sizes of $ v _t $, $ \mu _{t2} $ and $ v _i $. As $ v _i $ is fixed to be $ \sim 100 $ GeV and, as we will see later on, $ v _t $ is constrained to be $ \lsim O (1) $ GeV, we will always take $ v _t \ll v _i $. In this limit, the masses of $h$, $H$ and $H ^+$ are rather insensitive to $ v _t $ and $ \mu _{t2} $, while the masses of $H _t$, $A$, $H _t ^+$ and $H ^{++}$ strongly depend on them. As $\mu _{t2}$, in principle, remains as a free parameter, we can distinguish three different regimes according to its size: 

\begin{itemize}

\item  $ \mu _{t2} \sim v_t \ll v _i $: In this case, $ \mu _{t2}$ has little influence on the masses of $h$ and $H$. $H$ remains always heavier than $h$, with a mass around the $ 100-300 $ GeV range, for $ \lambda$'s of order $\sim 1$. The masses of $ H _t $, $ A $, $H _t ^+$ and $ H ^{++} $ are controlled by the ratio $\mu _{t2} / v _t$, with $ H _t $ and $ A $ nearly mass degenerate. In particular, for $\mu _{t2} = v _t$, the masses are around $200$~GeV. Such low masses can be dangerous in light of existing bounds \cite{Basso:2012st,Cheon:2012rh,Eberhardt:2013uba,Broggio:2014mna,vonBuddenbrock:2016rmr,Dery:2017axi,Basler:2017nzu,Patrick:2017ele,Ren:2017jbg}.

\item $ \mu _{t2} \sim v _i $: In this scenario the spectrum is shifted up and the scalar masses can be significantly larger than $100$~GeV. $A$, $H_t$, $H _t ^+$ and $ H ^{++} $ are mass degenerate and may reach masses in the TeV domain. For example, taking $ \mu _{t2} = v _2 = 100 $ GeV and $ v _t = 100 $ MeV, we obtain $ m _h = 125 $ GeV, $ m _H = 404 $ GeV, $ m _{H ^+} = 507 $ GeV, $ m _{H _t} \simeq m _A \simeq m _{H _t ^+} \simeq m _{H ^{++}} \simeq 2.65 $ TeV. We have adopted $ \lambda _1 = 1.6 $, $ \lambda _2 = 0.9 $, $ \lambda _3 = 7.7 $, $ \lambda _4 = - 8.5 $, $ \lambda _{ti} = \lambda _{tti} = 0.5 $.

\item $ \mu _{t2} \gg v_i$: This case can be recognized as the canonical type II seesaw scenario, in which $ m _t $ and $ \mu _{t2} $ come from new physics at very high energy scale, like Grand Unification scale. In this case, only $ h $, $ H $ and $ H ^+ $ remain in the weak scale, while $ H _t $, $ A $, $ H _t ^+ $ and $ H ^{++} $ decouple and are still degenerate, getting very high masses of order $ \sim v \sqrt{\mu _{t2} / v _t} $. \\

\end{itemize}

As stressed above, the masses of $ H _t $, $ A $, $ H _t ^+ $ and $ H ^{++} $ are always close to each other because their masses follow $ m _{H _i} ^2 = m _t ^2 + O (v, v _t) $, with $ m _t ^2 \simeq \mu _{t2} v _2 ^2 / \sqrt{2} v _t $. Thus, for small $v _t$, the $ m _t ^2 $ term is the dominant one, so that the masses are all approximately given by $ m _{H _i} ^2 \simeq m _t ^2 $. The mass splittings are controlled by the scalar VEVs. At leading order, 
\begin{equation}
\begin{split}
& m _{H _t} ^2 - m _A ^2 \simeq O (v _t ^2) , \\
m _A ^2 - m _{H ^+ _t} ^2 \simeq & \text{\ } m _{H ^+ _t} ^2 - m _{H ^{++}} ^2 \simeq \frac{1}{4} ( \lambda _{tt1} v _1 ^2 + \lambda _{tt2} v _2 ^2 ) .
\end{split}
\end{equation}
These mass splittings are noticeable only when the masses are small, i.e., for small $ \mu _{t2} $. For $ \mu _{t2} \gsim v _i $, they are basically mass degenerate. \\

As aforementioned, the masses of $h$, $H$ and $H ^+$ are less sensitive to $ v _t $ and $ \mu _{t2} $, and depend mostly on the VEVs $ v _i $ and the $ \lambda _i $'s. Therefore, they naturally lie at the weak scale. As for $H ^{+}$, we find $ m _{H ^+} ^2 \simeq - \frac{1}{2} \lambda _4 v ^2 $ (see eq. (\ref{masses_charged_app2_1}) in the Appendix \ref{sec:app2}), which requires $ \lambda _4 $ to be negative.
If we took $ | \lambda _4 | > 1 $, we would have charged scalar masses above $500$~GeV, which can easily evade existing limits \cite{Kakizaki:2003jk,Garayoa:2007fw,Mantry:2007ar,Chen:2012vm,delAguila:2013yaa,delAguila:2013mia,Blunier:2016peh,Boos:2018fnt}. In order to have scalar masses above 500GeV we need couplings larger than the unit. This typically true in seesaw type II models. \\

Returning to the mass expressions of the CP-even scalars, eqs. (\ref{mass_neutral_approximate_1}) and (\ref{mass_neutral_approximate_2}), we see that ifthere were just the two scalar doublets, the neutral scalar masses would be given by these expressions with $ v _t , \mu _{t2} $ set to zero. However, the scalar triplet generate negative correction terms proportional to $\mu _{t2} v _t $, so that these scalars become lighter than they would be if there was not the triplet. However, as shown in Figure~\ref{SM_like_Higgs_parameter_space}, the parameter space allows to fit a mass $m _h = 125$ GeV for the Higgs boson $h$. Furthermore, the eqs. (\ref{mass_neutral_approximate_1}) and (\ref{mass_neutral_approximate_2}) imply an upper bound in the combination $ \mu _{t2} v _t $: a large value of $ \mu _{t2} $ must be balanced by a small value of $ v _t $ in order to avoid negative squared masses for $h$ and $H$. Looking at these equations, we conclude that in order to preserve the masses positive, $ v _t $ must satisfy, 
\begin{equation}
\label{seesaw_relation_scalar_tachyon}
v _t \lsim \frac{( \lambda v ) ^2}{\mu _{t2}} ,
\end{equation} where we assume $\lambda v \sim 100$~GeV. For instance, $ \mu _{t2} \sim 10 ^{14} $ GeV implies $ v _t \lsim 10 ^{-1} $ eV. Thus, one may naturally generate small $ v _t $ taking $ \mu _{t2} $ at a Grand Unification scale \cite{Arhrib2011}. Notice that eq.~(\ref{seesaw_relation_scalar_tachyon}) is another kind of seesaw relation between $ v _t $ and $ \mu _{t2} $ valid in the limit $ v _t \ll \mu _{t2} $, which is independent of the relation in eq.~(\ref{seesaw_II_relation}). \\

Now that we have finished with the scalar sector, we will derive the masses of the gauge bosons. 

\subsection{Mass Spectrum - Gauge Bosons}

The Lagrangian for the kinetic terms of the gauge fields associated to the hypercharge $U(1)_Y$ and the $U(1)_X$ symmetry is given by,
\begin{equation}
\mathcal{L} _{\text{gauge}} = - \frac{1}{4} \hat{B} _{\mu \nu} \hat{B} ^{\mu \nu} + \frac{\epsilon}{2 \cos \theta _W} \hat{X} _{\mu \nu} \hat{B} ^{\mu \nu} - \frac{1}{4} \hat{X} _{\mu \nu} \hat{X} ^{\mu \nu},
\end{equation}
where $\epsilon$ is the kinetic mixing parameter. \\

A canonical gauge kinetic Lagrangian is obtained through a $ GL(2,R) $ rotation on the fields $ \hat{B}_\mu $ and $ \hat{X}_\mu $,
\begin{equation}
\begin{split}
\label{bosons_simet_aprox}
& \hat{X} _{\mu} \simeq  X _{\mu} \\
\hat{B} _{\mu} \simeq & B _{\mu} + \frac{\epsilon }{\cos \theta _W} X _{\mu} ,
\end{split}
\end{equation}
so that the covariant derivatives (\ref{covariant_derivative_1}) and (\ref{covariant_derivative_2}) become,

\begin{equation}
D _\mu \Phi _i = \partial _\mu \Phi _i + i g \tau ^a W _\mu ^a + i g ' \frac{Y}{2} B _\mu \Phi _i + \frac{i}{2} G _{X _i} X _\mu \Phi _i ,
\end{equation}
\begin{equation}
\label{der_cov_u1_diag}
D _\mu \Delta = \partial _\mu \Delta + i g [ T^a W_\mu ^a , \Delta ] + i g ' \frac{Y _t}{2} B _{\mu} \Delta + \frac{i}{2} G _{X _t} X_\mu \Delta ,
\end{equation}
where $ G _{Xi} = g ' \frac{\epsilon Y _i}{\cos \theta _W} + g _X Q _{X i} $ and $ G _{X _t} = g ' \frac{\epsilon Y _t}{\cos \theta _W} + g _X q _{X _t} $.\\

After spontaneous symmetry breaking and performing the electroweak rotation,
\begin{equation*}
\begin{split}
B_\mu & =  \cos \theta _W A_\mu - \sin \theta _W Z_\mu ^0,  \\
W_\mu ^3 & =  \sin \theta _W A_\mu + \cos \theta _W Z_\mu ^0 ,
\end{split}
\end{equation*}
the spectrum of vector bosons turns out to be comprised of: the charged $W_\mu^\pm$; the photon, $A_\mu$; two neutral states, $Z_\mu^0$ and $X_\mu$,  mixing to each other. They have the following mass Lagrangian 
\begin{equation}
\mathcal{L} _{\text{mass}} = m_W ^2 W_\mu ^- W ^{+ \mu} + \frac{1}{2} m _{Z ^0 X} ^2 Z_\mu ^0 Z^{0 \mu} - m_{Z^0X}^2 Z_\mu ^0 X ^\mu + \frac{1}{2} m_X ^2 X_\mu X ^\mu ,
\end{equation}
where, 
\begin{equation}
\label{w_boson_mass}
m_W ^2 = \frac{1}{4} g^2 ( v ^2 + 2 v _t ^2 ), 
\end{equation}
\begin{equation}
m_{Z^0} ^2 = \frac{1}{4} g_Z ^2 ( v ^2 + 4 v _t ^2 ),
\end{equation}
\begin{equation}
m _{Z ^0 X} ^2 = \frac{1}{4} g _Z ( G_{X1} v_1 ^2 + G_{X2} v_2 ^2 + 2 G _{X _t} v _t ^2 ),
\end{equation}
\begin{equation}
m_X ^2 = \frac{1}{4} ( v_1 ^2 G_{X1} ^2 + v_2 ^2 G_{X2} ^2 + G _{X _t} ^2 v _t ^2 ), ,
\end{equation}
with $ g _Z ^2 = g ^2 + g^{' 2} = g ^2 / \cos ^2 \theta _W $, $v ^2 = v _1 ^2 + v _2 ^2$ and $v ^2 + 2 v _t ^2 = ( 246 \text{GeV} ) ^2$. \\

We see that the $W_\mu ^\pm $ bosons are already the mass-eigenstates with mass $ m_W $. The $Z$ and $Z^\prime$ gauge bosons on the other hand mix and lead to the following mass matrix,

\begin{equation}
M_{Z ' Z} ^2 = \begin{pmatrix} m_{Z^0} ^2 & - m _{Z ^0 X} ^2 \\ - m _{Z ^0 X} ^2 & m_X ^2 \end{pmatrix}.
\end{equation}

The diagonalization leads to,

\begin{equation}
\begin{split}
\label{autovalores_matriz_zz}
m_{Z} ^2 &= \frac{1}{2} \left[ m_{Z ^0} ^2 + m_X ^2 + \sqrt{ \left( m_{Z ^0} ^2 - m_X^2 \right) ^2 + 4 \left( m _{Z ^0 X} ^2 \right) ^2 } \right], \\
m_{Z '} ^2 &= \frac{1}{2} \left[ m_{Z ^0} ^2 + m_X ^2 - \sqrt{ \left( m_{Z ^0} ^2 - m_X^2 \right) ^2 + 4 \left( m _{Z ^0 X} ^2 \right) ^2} \right] .
\end{split}
\end{equation} where, 

\begin{equation}
\label{rotacao_zz_fisicos}
\begin{pmatrix} Z_\mu \\ Z ' _\mu \end{pmatrix} = \begin{pmatrix} \cos \xi & - \sin \xi \\ \sin \xi & \cos \xi \end{pmatrix} \begin{pmatrix} Z^0 _\mu \\ X_\mu \end{pmatrix},
\end{equation}
with $ \xi $ given by,
\begin{equation}
\tan 2 \xi = \frac{2 m _{Z ^0 X} ^2}{m ^2 _{Z^0} - m ^2 _{X}} .
\end{equation}

This mixing angle is constrained to be very small by the LEP electroweak precision measurements on the $ Z $ boson pole \cite{Agashe:2014kda}. Thus, 
\begin{equation}
\xi \simeq \frac{m _{Z ^0 X} ^2}{m^2 _{Z^0} - m^2 _{X}} .
\end{equation}
Also, as we are interested in a light $ Z' $, we will assume the limit $ m ^2 _{Z^0} \gg  m ^2 _{X} $ (which implies $m ^2 _{Z^0} \gg  m _{Z ^0 X} ^2$ as well). In this limit,
\begin{equation}
\label{xi_delta}
\xi \simeq \frac{m _{Z ^0 X} ^2}{m^2 _{Z^0}} ,
\end{equation}and we can write approximate expressions for the masses of $ Z $ and $ Z' $ as

\begin{equation*}
\begin{split}
m _{Z , Z '} ^2 
& = \frac{1}{2} \left\{ m_{Z ^0} ^2 + m_X ^2 \pm \left( m_{Z ^0} ^2 - m_X^2 \right) \left[ 1 + \frac{4 \left( m _{Z ^0 X} ^2 \right) ^2}{\left( m_{Z ^0} ^2 - m_X^2 \right) ^2} \right] ^{\frac{1}{2}} \right\} \\
&\simeq \frac{1}{2} \left[ m_{Z ^0} ^2 + m_X ^2 \pm \left( m_{Z ^0} ^2 - m_X^2 + \frac{2 \left( m _{Z ^0 X} ^2 \right) ^2}{m_{Z ^0} ^2} \right) \right] .
\end{split}
\end{equation*}\\

For $ Z $, we have
\begin{equation*}
m _Z ^2 \simeq m_{Z ^0} ^2 + \frac{\left( m _{Z ^0 X} ^2 \right) ^2}{m_{Z ^0} ^2} ,
\end{equation*}and, at leading order, $ m _Z ^2 \simeq m_{Z ^0} ^2 $,
\begin{equation}
\label{z_boson_mass}
m_{Z} ^2 \simeq \frac{1}{4} g _Z ^2 ( v ^2 + 4 v _t ^2 ) .
\end{equation}
For $ Z ' $,
\begin{equation*}
\begin{split}
m_{Z '} ^2 
& \simeq m_X ^2 - \frac{\left( m _{Z ^0 X} ^2 \right) ^2}{m_{Z ^0} ^2} \\
& \simeq \frac{g _X ^2}{4} ( Q _{X _1} - Q _{X _2} ) ^2 \frac{v _1 ^2 v _2 ^2}{v ^2} ( 1 - \frac{4 v _t ^2}{v ^2}) .
\end{split}
\end{equation*}
In terms of $ \beta $, defined by $ \tan \beta = v _2 / v _1 $ (see Appendix \ref{sec:app2}),
\begin{equation}
m_{Z '} ^2 \simeq \frac{g _X ^2}{4} ( Q _{X _1} - Q _{X _2} ) ^2 v ^2 \sin ^2 \beta \cos ^2 \beta ( 1 - \frac{4 v _t ^2}{v ^2} ) .
\label{eqMZprime}
\end{equation}
Note that the presence of the triplet induces only a tiny correction proportional to $ (v_t / v) ^2 $, so that the addition of a triplet scalar cannot generate a heavy $Z^\prime$, as opposed to the singlet case \cite{Campos:2017dgc}. Thus, the $ Z' $ mass lies below the electroweak scale, being controlled by the value of $ g _X $. For instance, taking $\tan \beta = 10 $ and $Q _{X1} - Q _{X2} = 1$, $m _{Z'}$ varies from $1 \text{MeV} - 1 \text{GeV}$, for $g _X$ in the range of $10 ^{-3} - 10 ^{-1}$, regardless of the value of $v _t$, as long as $v _t < 2 \text{\ GeV}$. 

In summary, we  have proposed a type II seesaw mechanism for neutrino masses within the scope of 2HDM which prevent FCNI via gauge symmetries. Having discussed the mass spectrum of the model, we now will pay attention to some phenomenological constraints.

\section{Phenomenological constraints}
\label{pheno}

\subsection{Electroweak Precision}
The $ \rho $ parameter,
\begin{equation}
\rho = \frac{m _W ^2}{m _Z ^2 \cos ^2 \theta _W} ,
\end{equation}
which measures the relative intensity between the neutral and charged currents, is very accurately determined experimentally, $ \rho = 1.00039 \pm 0.00019 $ \cite{Tanabashi:2018oca} at $1\sigma$ level. In the SM, the $ \rho $ parameter is equal to $ 1 $ at tree level, and its good agreement with the experimental value poses tight constraints on new physics models with extended scalar sector. In our model, the $ \rho $ parameter  places an upper bound on the VEV of the triplet scalar because it contributes to the masses of $Z$ and $W ^\pm$ bosons according to eqs. (\ref{w_boson_mass}) and (\ref{z_boson_mass}),  translating into 
\begin{equation}
\rho = \frac{v ^2 + 2 v _t ^2}{v ^2 + 4 v _t ^2} .
\end{equation}

Hence at $3\sigma$  we obtain
\begin{equation}
v _t \leq 2.3 \text{\ GeV} ,
\end{equation}
where we used $ v ^2 + 2 v_t ^2 = 246 ^2\, \text{GeV} ^2$. As we are interested in a small $ v _t $ for the generation of tiny neutrino masses, this constraint can be easily satisfied in our model. Notice that as $v_t$ becomes very small the scalar masses increase as can be seen, for instance in eq.(\ref{eqpseudoA}) and eq.(\ref{eqdoubly}).

\subsection{Collider Bounds}

\subsubsection{LHC - $Z^\prime$}
The $ U(1) _X $ symmetry is spontaneously broken by the VEV of the doublets and the triplet, which also contributed to the mass generation of the $Z ' $ vector boson. As $ v _t $ is small and $ v $ is at the electroweak scale, the $ Z ' $ mass will be at the electroweak scale or below, depending on the value of $ g _X $ and the other parameter such as  $ \tan \beta $. Such a light $Z^\prime$ is subject to a variety of experimental constraints. Notice that we have two possible $Z^\prime$ models ($U(1)_N$ or $U(1)_{Y^\prime} $), one which resembles the sequential $Z^\prime$ model, and other the dark photon model. Concerning the latter, LHC bounds are weakened because the $Z^\prime-Z$ mixing is necessarily small, and that would suppress its production cross section at the LHC \cite{Biswas:2016jsh,Barello:2016zlb,CMS:2018lqx}. \\

As for the $U(1)_{Y^\prime}$ model, we do not have much freedom since the SM fermions are charged under $U(1)_{Y^\prime}$ the production cross section is much larger. In this scenario the LHC bounds are rather restrictive. Assuming $g_X=1$ the LHC severely rules $Z^\prime$ masses below $3$~TeV \cite{Aaboud:2018bun}. In our model the $Z^\prime$ mass is set by $g_X$. In order to have $Z^\prime$ masses around $100$~GeV, $g_X$ should be around $0.1$, which is not sufficiently small to evade LHC limits \cite{Allanach:2015gkd}. If we adopt $g_X=0.01$ we will get $m_{Z^\prime} = 1$~GeV, and for such small coupling we can easily evade LHC limits \cite{Allanach:2015gkd}. We have used eq.~(\ref{eqMZprime}) and assumed  $Q_{X1}$ of the same order of $Q_{X2}$ to find the corresponding $Z^\prime$ mass. We point out that the kinetic mixing parameter $\epsilon$ while not relevant for the $Z^\prime$ mass it is important to determine the $Z^\prime$ interactions with SM fermions. The conclusions drawn above are valid for sufficiently small kinetic mixing.\\ 

\subsubsection{LHC - Doubly Charged Scalar}
Regarding the scalar spectrum of our model, the most relevant ones come from LHC searches for heavy Higgs and triplet scalars. The cleanest signature signal is the doubly charged Higgs. We have then implemented the model in Madgraph \cite{Alwall:2007st,Alwall:2011uj} and followed the recipe described in \cite{CMS:2017pet}. Assuming no hierarchy in the Yukawa couplings the doubly charged scalar decays essentially, with equal branching ratios, into charged leptons. That said, we found the current LHC bound with $\mathcal{L}=36fb^{-1}$ of integrated luminosity and performed future projects for the High Luminosity and High Energy LHC setups as summarized in the Table \ref{tab:doubly}.  \\

\begin{table}[!h]
    \centering
    
    \begin{tabular}{|c|c|}
    \hline
 LHC 13TeV - $\mathcal{L}= 12.9 fb^{-1}$&  $m_{H^{++}} > 760$~GeV  \\
 LHC 13TeV - $\mathcal{L}= 36 fb^{-1}$& $m_{H^{++}} > 980$~GeV  \\
High-Luminosity LHC - $\mathcal{L}= 1000 fb^{-1}$ & $m_{H^{++}} > 1.9$~TeV\\
High-Energy LHC 27TeV, $\mathcal{L}= 1000 fb^{-1}$ & $m_{H^{++}} > 3$~TeV, \\
           \hline
    \end{tabular}
    \caption{Summary of collider bounds on the doubly charged scalar in our model using current and planned configurations. We used 13TeV of center-of-mass energy for the LHC configurations, whereas 27TeV for  the high-energy upgrade. We can see that LHC and its upgrade will be paramount to probe the model up to the TeV scale. }
    \label{tab:doubly}
\end{table}

In the light of current bounds our model is in agreement with existing bounds if we take  $\mu_{t2} \leq v_i$ which predicts masses at the TeV scale as discussed previously. We highlight that we need couplings larger than one to find charged scalar masses above the TeV scale. Therefore, LHC and its planned upgrade will be important because it will probe a large portion of the model. The presence of a doubly charged scalar is the key signature of the type II seesaw mechanism. 

We highlight that these bounds reply on lepton flavor violation channels with a degenerate neutrino mass spectrum with absolute mass around $0.1$~eV \cite{CMS:2017pet}. The consideration of different mass hiearchies will not bring much impact to our paper which focuses on the proposition of a new 2HDM model. Anyways, strictly speaking, one should keep in mind that different mass hierarchies are subject to different collider bounds as pointed out \cite{Ferreira:2019qpf}. Anyway, if we find ourselves in a situation where both doubly charged scalar and $Z^\prime$ fields are observed, our model stands as a potential well motivated environment for them. 

\subsection{LHC- Higgs}

Now the higgs decays to SM fermions and gauge bosons fermions have been constrained \cite{Almeida:2018cld,Alves:2018nof,Cepeda:2019klc}, one can use Higgs data to place important limits on the model.

The couplings of the Higgs-like scalar $h$ with the SM fermions and gauge bosons are given by,
\begin{equation}
\mathcal{C} _{h \bar{f} f} = \frac{ ( s _\alpha c _{\alpha _2} - c _\alpha s _{\alpha _1} s _{\alpha _2} ) }{s _\beta} \mathcal{C} _{h \bar{f} f} ^{SM}
\end{equation}
\begin{equation}
\mathcal{C} _{h W W} = ( c _\alpha c _{\alpha _1} c _\beta + s _\alpha c _{\alpha _2} s _\beta - c _\alpha s _{\alpha _1} s _{\alpha _2} s _\beta ) \mathcal{C} _{h W W} ^{SM}
\end{equation}
\begin{equation}
\mathcal{C} _{h Z Z} = ( c _\alpha c _{\alpha _1} c _\beta + s _\alpha c _{\alpha _2} s _\beta - c _\alpha s _{\alpha _1} s _{\alpha _2} s _\beta ) \mathcal{C} _{h Z Z} ^{SM} ,
\end{equation}
where $\mathcal{C} _{h \bar{f} f} ^{SM} = \frac{m _f}{v}$, $\mathcal{C} _{h W W} ^{SM} = \frac{1}{2} g ^2 v$ and $\mathcal{C} _{h Z Z} ^{SM} = \frac{1}{2} \frac{g ^2 v}{\cos ^2 \theta _W}$. In the expressions for the gauge bosons we have neglected small terms proportional to $\epsilon$ and $\sin \xi$.
As shown in the Appendix \ref{sec:app1}, the angles $\alpha _1$ and $\alpha _2$ (and also $\beta _1$ and $\beta _2$) are suppressed by $v _t / v _2$. Then taking $\alpha _1 , \alpha _2 \rightarrow 0$ in the above expressions, we get,
\begin{equation}
\mathcal{C} _{h \bar{f} f} = \frac{s _\alpha}{s _\beta} \mathcal{C} _{h \bar{f} f} ^{SM}
\end{equation}
\begin{equation}
\mathcal{C} _{h W W} = c _{\beta - \alpha} \mathcal{C} _{h W W} ^{SM}
\end{equation}
\begin{equation}
\mathcal{C} _{h Z Z} = c _{\beta - \alpha} \mathcal{C} _{h Z Z} ^{SM} ,
\end{equation}
with $ c _{\beta - \alpha} \equiv \cos (\beta - \alpha) $. When $\alpha = \beta$, we fall in the alignment limit \cite{PhysRevD.67.075019,Carena2014}. In this regime, $h$ couples to the SM particles identically to the SM Higgs. Conversely, the couplings of the havier Higgses $H$ and $H _t$, which are proportional to $s _{\beta - \alpha}$ and $s _{\alpha _i}$, respectively, vanish in this limit.

For $ c _{\beta - \alpha} \sim 1 $, $\tan \beta$ can take on essentially any value, as long as the Higgs-like decays are concerned (see, e.g., Fig. 3 of Ref. \cite{Campos:2017dgc}\footnote{Care must be taken when comparing our results with the ones in Ref.\cite{Campos:2017dgc}, because the physical scalars and mixing angle $\alpha$ are defined following different conventions (see Eq. (4.13) from that paper and compare with Eq. (\ref{rot_to_phy_basis_cp_even}) in Sec. 3.1.). For this reason, $c _{\beta - \alpha} = 0$ in Fig.3 of that paper, is equivalent to $c _{\beta - \alpha} = 1$ in ours.} ). Regarding the charged Higgs $H ^+$, its coupling to fermions is suppressed by a factor of $\tan \beta$. Therefore, large values of $\tan \beta$ weakens the LHC limits. In summary, our model can be made fully consistent in the alignment limit with no prejudice. \\ 


\subsection{LHC- Heavy Higgs}

An interesting signature of our model is the decay of the heavy Higgs, H, into heavy gauge bosons \cite{Ko:2014uka},

\begin{equation}
\Gamma (H\rightarrow Z'Z')=\frac{g^2}{128\pi}\frac{m_H^2}{m^2_Z}(\delta \tan\beta)^4\left(\frac{\cos^3\beta\cos\alpha-\sin^3\beta \sin\alpha}{\cos\beta\sin\beta} \right)^2
\end{equation}where,

\begin{equation}
\delta=g_{X}\frac{\cos\theta_W m_Z}{g m_{Z^\prime}}\left(Q_{X1}\cos^2\beta + Q_{X2}\sin^2\beta\right).
\end{equation}

This decay is kinematically available because the $Z^\prime$ gauge boson is very light. Depending on the magnitude of $g_X$, $Z^\prime$ might decay inside the detector. Thus, the possible signature of this heavy scalar is the four lepton channel \cite{Sirunyan:2017lae,CMS:2017uzk}. We plan to investigate the LHC discovery reach of this decay mode in the foreseeable future. A detail phenomenology is out of the scope of the current paper, but it is important to stress that in the aforementioned alignment limit, this decay channels closes and the bounds stemming from heavy Higgs weaken \cite{Ko:2013zsa}.

\subsubsection{Belle-II and KLOE2}

Belle and KLOE collaborations represent $e^+e^-$ colliders searching for light gauge bosons with the $\epsilon/2 F^{\mu\nu} F_{\mu\nu}^\prime$. 
The two models proposed here feature a similar term. In the $U(1)_N$ model SM fermions are uncharged under $U(1)_N$, thus the $Z^\prime$ will couple to SM fermions only via its mixing with the Z boson generated by the presence of the kinetic mixing.  In this case, our model would a UV complete version of the simplified dark photon model \cite{Fayet:1990wx,Fayet:2007ua}. This scenario for heavy $Z^\prime$ masses was investigated in \cite{Arcadi:2018tly}.  For the $U(1)_{Y^\prime}$ model, where the SM fermions are charged under the gauge symmetry, if we take $g_X \ll 1$ and $g_X < \epsilon$, again the model falls back to the dark photon model because the $Z^\prime$ interactions to SM via the kinetic mixing would more pronounced, the experimental limits on dark photon become applicable to our study. \\

\begin{figure}
    \centering
    \includegraphics[scale=0.7]{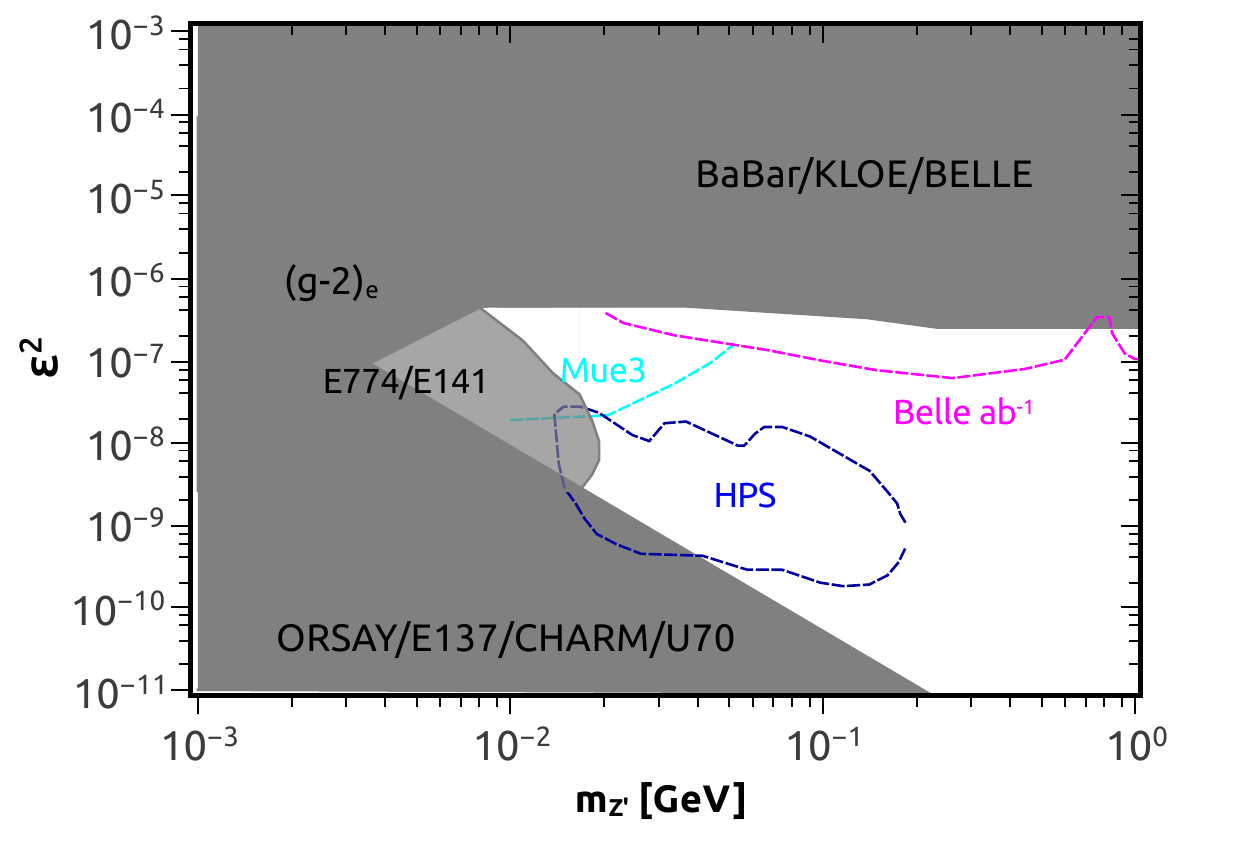}
    \caption{Summary of experimental limits applicable to the model $U(1)_N$ and to the model $U(1)_{Y^\prime}$ assuming $g_X \ll \epsilon$. Current limits are in gray while projected ones in color. }
    \label{fig:darkphoton}
\end{figure}

In summary, the experimental limits derived for dark photon models apply here, except in the case where $g_X \gg \epsilon$ and $m_{Z^\prime} \gg 1$~GeV. Experimental collaborations usually display their bounds in terms of $\epsilon^2$. In Figure~\ref{fig:darkphoton} we display a summary of the existing (gray) and planned (color) constraints.  For $m_{Z^\prime} \sim 10-30$~MeV, current bound impose $\epsilon < 10^{-4}$, limiting the region of parameter of our model. Anyhow, we emphasize that we can still obey such bounds by taking $g_X$ and $\epsilon$ to be very small as it is usually assumed in dark photon models.

\subsection{Accelerators}

There are several accelerators using electron/positron or hadronic beams which search for bremsstrahlung of dark photons or its appearance in meson decays. These bounds are inside the gray region in Figure~\ref{fig:darkphoton}. It is important to point out the future sensitivity on the flavor violating decay $\mu \rightarrow 3e$ \cite{Blondel:2013ia}, in case of no signal, will give rise to the upper limit in cyan. Moreover, The Heavy Photon Search Experiment (HPS) which was already installed at SLAC collides highly energetic electrons into a tungsten target, and in the process electrons may radiate dark photons. The experimental sensitivity of HPS is shown in blue.

\subsection{Low Energy Probes}

The muon anomalous magnetic moment (g-2) \cite{Lindner:2016bgg}, neutrino-electron scattering \cite{Lindner:2018kjo} and atomic parity violation \cite{Campos:2017dgc} provide complementary but subdominant limits to our model. For instance, neutrino-electron scattering rules out $\epsilon  > 10^{-5}$ \cite{Lindner:2018kjo}. One cannot accommodate g-2 with the $U(1)_{Y^\prime}$ model because the electrons are charged under the gauge symmetry, and these couplings are subject to tight constraints \cite{Freitas:2014pua,Kaneta:2016uyt,Kowalska:2017iqv}. In the $U(1)_N$ model, where our model resembles the dark photon one, the favored region to explain g-2 has already been excluded \cite{Lindner:2016bgg}.

\subsection{Dark Matter Possibility}

In the two models described here, there are no dark matter candidates. One could simply add a vector-like fermion charged under the gauge symmetry while preserving the gauge anomaly cancellation.  The dark matter relic abundance, direct detection and indirect detection signals would be governed the kinetic mixing term and the $g_X$ gauge coupling. For concreteness, if we take the $U(1)_N$ model, where SM fermions are not charged under the new gauge symmetry, the dark matter phenomenology would be similar to the dark photon portal investigated recently in the literature \cite{Dutra:2018gmv}. It has been shown that in this setup if the dark matter mass is smaller than the $Z^\prime$ one, only s-channel processes would be relevant, and this case is nearly ruled out by current data for dark matter masses above $10$~MeV. There is tiny region for $\epsilon \sim 10^{-5}$ and $m_{Z^\prime} \sim 10$~MeV which obeys on existing limits and reproduce the correct relic density. If the dark matter particle is heavier than the $Z^\prime$ then the secluded dark matter setup arises \cite{Profumo:2017obk}, scenario which is much less restricted by data. In the near future we plan to carry out a detailed dark matter phenomenology in both models taking into account the particularities such as the presence of both mass mixing and kinetic mixing terms as well as the existence of many other scalars which might alter the overall predictions. 

\section{Discussion}

There are important things to be stressed about the two  models we proposed to explain neutrino masses and the absence of flavor changing interactions:

\begin{itemize}

\item The gauge symmetry imposed to distinguish $\Phi_1$ from $\Phi_2$ and then allow just one scalar doublet to couple to SM fermions gives rise to two very different type of models. In the $U(1)_N$ model, the SM fermions are uncharged under the gauge symmetry, and the corresponding massive $Z^\prime$ only couples to SM fermions via its mixing with the Z boson. In the $U(1)_{Y^\prime}$ setup, the $Z^\prime$ gauge boson will have a neutral current with SM fermions determined by the $U(1)_{Y^\prime}$ SM fermion charges, leading to a sequential $Z^\prime$ model \cite{Alves:2015pea}. 

\item Since we have added a triplet scalar to explain neutrino masses via a type II seesaw mechanism, nothing prohibits from one to consider off-diagonal yukawa couplings, involving for instance the scalar triplet,  to be non-vanishing. This will lead again to flavor changing interactions and give rise to $\mu \rightarrow e\gamma$, $\mu \rightarrow 3e$, $\mu-e$ conversion processes which are rather restricted by data. In particular, the product of the yukawa terms are limited to be smaller than $10^{-7}$ \cite{Lindner:2016bgg}. Anyways, this feature is common in the models which extend the SM scalar sector and, thus, we can set to zero off-diagonal couplings involving the extra scalars without prejudice. 

\item The additional gauge symmetry allow us to easily introduce a dark matter candidate, a vector-like fermion, without spoiling the anomaly cancellation requirements. In summary, we argue that the addition of a gauge symmetry and a triplet scalar is well-motivated since it adds nice features to the original 2HDM proposal.

\item The bounds discussed previously can be safely satisfied by taking $g_X$ and $\epsilon$ to be sufficiently small, below $10^{-3}$ similarly to dark photon models. 

\end{itemize}

\section{Conclusions}
\label{sec:concl}

Two Higgs Doublet Models represent interesting extensions of the Standard Model. However they lack neutrino masses and a first principle explanation for the absence of flavor changing neutral interactions. Typically, a $Z_2$ symmetry is invoked where one scalar doublet is charged under, to prevent the other to couple to Standard Model fermions. 

In this work, we proposed two models where neutrino masses are explained within the type II seesaw mechanism via the addition of a scalar triplet and a gauge symmetry that allows only one scalar doublet to couple to fermions. In this way can simultaneously explain neutrino masses and  avoid flavor changing neutral interactions.  

We have investigated several constraints coming from low energy probes, electroweak precision and collider. In particular, we derived  collider bounds on the mass of the doubly charged scalar using current and planned LHC reach with high-luminosity and high-energy configurations. We discussed which regions of parameter space are consistent with current data to  conclude that both models stand as viable alternatives to the original Two Higgs Doublet Model proposal. 

\clearpage

\section*{Acknowledgments}

The authors are grateful to Carlos Yaguna, Carlos Pires and Diego Cogollo for their comments. D.A.C. acknowledges support from UFRN and MEC while T.B.M.  thanks CAPES for financial support. A.G.D. is partially supported by the Conselho Nacional de Desenvolvimento Cient\'{\i}fico e Tecnol\'ogico (CNPq) under the  grant 306636/2016-6.  F.S.Q. is partially financed by ICTP-SAIFR grant 2016, UFRN and MEC. A.G.D. is grateful to UFRN for the hospitality during the early stages of this project. 

\appendix

\section{Vacuum Stability Bounds } 
\label{sec:appvcb}

In this Appendix we give the vacuum stability bounds, also known as bounded from below conditions, for the scalar potentials in the sections \ref{typeImodel} and \ref{typeIImodel}. General discussions concerning vacuum stability bounds for a scalar potential of few scalar fields can be found, for example, in Refs. \cite{Kannike:2012pe,Kannike:2016fmd}.  In order to determine the vacuum stability bounds it is sufficient to consider only the quartic terms in the potential, once they turn out to be the dominant contribution for large values of the fields. 

\subsection{Vacuum stability bounds for the 2HDM-U(1) with type I seesaw mechanism  }
\label{apps}

The vacuum stability bounds for the model with the two scalar doublets and a complex scalar singlet of the model with type I seesaw mechanism in section \ref{typeImodel} can be found following the same steps of Ref. \cite{Arhrib:2018qmw}, where it was treated the case with a real singlet. The analysis will be also useful next when dealing with the potential of the type II seesaw mechanism. 

It is convenient to parameterize the fields according to 
\begin{equation}
\begin{split}
r^2 & =  \Phi_1^\dagger \Phi _1 + \Phi_2^\dagger \Phi _2  + \Phi_s^\dagger\Phi_s,   \\
\Phi_1^\dagger \Phi _1 & =  r^2\sin^2\theta\cos^2\phi,\\
\Phi_2^\dagger \Phi _2 & =  r^2\sin^2\theta\sin^2\phi,\\
\Phi_s^\dagger \Phi _s & =  r^2\cos^2\theta, 
\end{split}
\label{param}
\end{equation}
in which $ 0\leq r\leq \infty$, $0\leq \theta\leq \frac{\pi}{2}$ and  $0\leq \phi\leq \frac{\pi}{2}$. We also define the ratio 
\begin{equation}
    \frac{\Phi _1 ^\dagger \Phi _2 }{|\Phi _1 || \Phi _2| }=c\,e^{i\alpha },
\end{equation}
with $0\leq c\leq 1$ and $\alpha\in \Re $. Thus, the potential of the quartic terms, $V_4$, from eqs. (\ref{potential_doblets_u1_x}) and (\ref{potential_doblets_singlet_u1_x}) is such that 
\begin{equation}
\begin{split}
 V_4/r^4 &=\left\{ \frac{\lambda_{1}}{2}\left(1-x\right)^{2}+\frac{\lambda_{2}}{2}x^{2}+\left(\lambda_{3}+\lambda_{4}c^{2}\right)x\left(1-x\right)\right\} y^{2}\\
 &+\frac{\lambda_{s}}{2}\left(1-y\right)^{2} +\left\{ \lambda_{s1}\left(1-x\right)+\lambda_{s2}x\right\} y\left(1-y\right)\\
 &= A_x y^{2}+B_x \left(1-y\right)^{2} + C_x y\left(1-y\right),
\end{split}
\label{vr4}
\end{equation}
where we define $x=\sin^2\phi$, $y=\sin^2\theta$ and the function $A_x$, $B_x$, $C_x$ which can be read from the first and second lines of eq. (\ref{vr4}). The condition $V_4>0$ implies that 
\begin{equation}
    \begin{split}
        \label{ABC}
        A_x>0,\hspace{0.4 cm}B_x>0,\hspace{0.2 cm} {\rm and} 
        \hspace{0.2 cm} 2\sqrt{A_x B_x}+C_x>0.
    \end{split}
\end{equation}
The first condition above, 
\begin{equation}
\label{Ax0}
\begin{split}
 A_x= \frac{\lambda_{1}}{2}\left(1-x\right)^{2}+\frac{\lambda_{2}}{2}x^{2}+\left(\lambda_{3}+\lambda_{4}c^{2}\right)x\left(1-x\right)>0, 
\end{split}
\end{equation}
is the requirement that the potential is  positive along the $y=1$ direction for large of the fields. It has the same form of eq. (\ref{vr4}) so that  the directions $x=0,\,1$ give
\begin{equation}
\label{l1l2s}
    \begin{split}
        \lambda_{1}>0,\hspace{0.5 cm} \lambda_{2}>0. 
    \end{split}
\end{equation}
Analogously, a condition similar to the last one in eq. (\ref{Ax0}) gives, for $c=0,\,1$,  
\begin{equation}
    \label{l34s}
    \begin{split}
        \lambda_{3}+\sqrt{\lambda_{1}\lambda_{2}}>0,\hspace{0.5 cm} \lambda_{3}+\lambda_{4}+\sqrt{\lambda_{1}\lambda_{2}}>0. 
    \end{split}
\end{equation}
The second condition in eq. (\ref{ABC}) is the requirement that the potential is  positive along the $y=0$ direction for large of the fields and gives
\begin{equation}
    \lambda_{s}>0.
\end{equation}
The last inequality in eq. (\ref{ABC}) arises from requiring that the discriminant of eq. (\ref{vr4}), as a polynomial in $y$, is negative, which implies that $4A_x B_x-C_x^{2}>0$. There are two cases to consider for $C_x$. One is the case in which $\lambda_{s1}>0$ and $\lambda_{s2}>0$ resulting in $C_x>0$ that automatically satisfies eq. (\ref{ABC}). The other case is for the couplings $\lambda_{s1}<0$ or/and $\lambda_{s2}<0$  so that we must have from the discriminant that
\begin{equation}
\label{disc}
    \begin{split}
      \left(\lambda_{1}\lambda_{s}-\lambda_{s1}^{2}\right)\left(1-x\right)^{2}+\left(\lambda_{2}\lambda_{s}-\lambda_{s2}^{2}\right)x^{2} +2\left\{ \left(\lambda_{3}+\lambda_{4}c^{2}\right)\lambda_{s}-\lambda_{s1}\lambda_{s2}\right\} x\left(1-x\right)>0.
    \end{split}
\end{equation}
In the same way that follows from eq. (\ref{Ax0}), this leads to the conditions 
\begin{equation}
    \label{ls1ls2} 
    \begin{split}
     & \lambda_{s1}+\sqrt{\lambda_{1}\lambda_{s}}>0\,,\hspace{0.8 cm}  \lambda_{s2}+\sqrt{\lambda_{2}\lambda_{s}}>0\,,\\
     & 2(\lambda_{3}\lambda_{s}-\lambda_{s1}\lambda_{s2}) +\sqrt{(\lambda_{1}\lambda_{s}-\lambda_{s1}^{2})(\lambda_{2}\lambda_{s}-\lambda_{s2}^{2})}>0\,,\\
     & 2(\lambda_{3}\lambda_{s}+\lambda_{4}\lambda_{s}-\lambda_{s1}\lambda_{s2}) +\sqrt{(\lambda_{1}\lambda_{s}-\lambda_{s1}^{2})(\lambda_{2}\lambda_{s}-\lambda_{s2}^{2})}>0.
    \end{split}
\end{equation}

\subsection{Vacuum stability bounds for the 2HDM-U(1) with type II seesaw mechanism  }
\label{appt}
In order to obtain the vacuum stability bounds for the potential with two scalar doublets plus a scalar triplet of the 2HDM-U(1) with type II seesaw mechanism in section \ref{typeIImodel}, we use the same sort of  parameterization as in  \ref{apps} 
\begin{equation}
\begin{split}
r^2 & =  \Phi_1^\dagger \Phi _1 + \Phi_2^\dagger \Phi _2  + {\rm Tr}(\Delta^\dagger\Delta) ,   \\
\Phi_1^\dagger \Phi _1 & =  r^2\sin^2\theta\cos^2\phi=r^2y(1-x),\\
\Phi_2^\dagger \Phi _2 & =  r^2\sin^2\theta\sin^2\phi=r^2y\,x,\\
 {\rm Tr}(\Delta^\dagger\Delta)  & =  r^2\cos^2\theta=r^2(1-y),  
\end{split}
\label{paramt}
\end{equation}
in which $ 0\leq r\leq \infty$, $0\leq \theta\leq \frac{\pi}{2}$ and  $0\leq \phi\leq \frac{\pi}{2}$. Now we define the ratios  
\begin{equation}
\label{ratri}
    \begin{split}
         \frac{\Phi _1 ^\dagger \Phi _2 }{|\Phi _1 || \Phi _2| }=c\,e^{i\alpha },\hspace{0.4 cm} \frac{\Phi_i^\dagger\Delta\Delta^\dagger \Phi _i}{\Phi_i^\dagger \Phi _i{\rm Tr}(\Delta^\dagger\Delta)}=\xi_i,\hspace{0.4 cm} \frac{{\rm Tr}(\Delta^\dagger\Delta)^2}{\left[{\rm Tr}(\Delta^\dagger\Delta)\right]^2}=\zeta,
    \end{split}
\end{equation}
with $0\leq c\leq 1$, $\alpha\in \Re $, $0\leq\xi_{1,2}\leq 1$ and $\frac{1}{2}\leq\zeta\leq 1$\footnote{The range of variation of $\xi_i$ is defined by the Cauchy-Schwarz inequality, while the one for $\zeta$ is obtained through minimization of $\frac{{\rm Tr}(\Delta^\dagger\Delta)^2}{\left[{\rm Tr}(\Delta^\dagger\Delta)\right]^2}=\frac{1+\beta^2}{(1+\beta)^2}$.}. Thus, the potential of the quartic terms, $V_{4t}$, from eq. (\ref{scalarpotential}) is such that 
\begin{equation}
\begin{split}
 V_{4t}/r^4 = A_x y^{2}+B_x \left(1-y\right)^{2} + C_x y\left(1-y\right).
\end{split}
\label{vr4t}
\end{equation}
where
\begin{equation}
    \label{ABCt}
    \begin{split}
        A_x & = \frac{\lambda_{1}}{2}\left(1-x\right)^{2}+\frac{\lambda_{2}}{2}x^{2}+\left(\lambda_{3}+\lambda_{4}c^{2}\right)x\left(1-x\right),\\
        B_x & =\frac{1}{2}\overline{\lambda_{t}}(\zeta) = \frac{1}{2}(\lambda_{t}+\lambda_{tt}\zeta),\\
        C_x & = \overline{\lambda_{t1}}(\xi_1)\left(1-x\right)+\overline{\lambda_{t2}}(\xi_2)x= (\lambda_{t1}+\lambda_{tt1}\xi_1)\left(1-x\right)+(\lambda_{t2}+\lambda_{tt2}\xi_2)x,
    \end{split}
\end{equation}
with $\overline{\lambda_{t}}(\zeta)$, $\overline{\lambda_{t1}}(\xi_1)$ and $\overline{\lambda_{t2}}(\xi_2)$ defined just for convenience. 

The direction $y=1$ implies that $A_x>0$ and we have the same conditions of eqs. (\ref{l1l2s}) and (\ref{l34s}), i. e., 
\begin{equation}
\label{l1l2l3l4t}
    \begin{split}
       & \lambda_{1}>0,\hspace{0.5 cm} \lambda_{2}>0,\\
       & \lambda_{3}+\sqrt{\lambda_{1}\lambda_{2}}>0,\hspace{0.5 cm} \lambda_{3}+\lambda_{4}+\sqrt{\lambda_{1}\lambda_{2}}>0. 
    \end{split}
\end{equation}

The direction $y=0$ in eq. (\ref{vr4t}) implies that $B_x>0$ and this gives the conditions, for the values $\zeta=0,\,\frac{1}{2}$, 
\begin{equation}
    \label{ltltt}
    \begin{split}
        \lambda_{t}>0,\hspace{0.5 cm} \lambda_{t}+\frac{\lambda_{tt}}{2}>0.
    \end{split}
\end{equation}

It remains to analyse the inequality $C_x+2\sqrt{A_x B_x}>0$, which guarantees that $V_{4t}>0$ given that $A_x>0$ and $B_x>0$. In the same way as discussed in the subsection \ref{apps}, there are two cases to consider for $C_x$. The case in which  $\overline{\lambda_{t1}}\left(\xi_{1}\right)>0$ and $\overline{\lambda_{t2}}\left(\xi_{2}\right)>0$ gives $C_x>0$, so that the inequality is automatically satisfied once $A_x B_x>0$. The other case is the one in which $\overline{\lambda_{t1}}(\xi_1)<0$ or/and $\overline{\lambda_{t2}}(\xi_2)<0$, and the condition from the negative discriminant from eq. (\ref{vr4t}) is 
\begin{equation}
    \label{disct}
    \begin{split}
       & \left(\lambda_{1}\overline{\lambda_{t}}\left(\zeta\right)-\overline{\lambda_{t1}}^{2}\left(\xi_{1}\right)\right)\left(1-x\right)^{2}+\left(\lambda_{2}\overline{\lambda_{t}}\left(\zeta\right)-\overline{\lambda_{t2}}^{2}\left(\xi_{2}\right)\right)x^{2}\\
 & +2\left(\left(\lambda_{3}+\lambda_{4}c^{2}\right)\overline{\lambda_{t}}\left(\zeta\right)-\overline{\lambda_{t1}}\left(\xi_{1}\right)\overline{\lambda_{t2}}\left(\xi_{2}\right)\right)x\left(1-x\right)>0.
    \end{split}
\end{equation}
This implies, from  $x=0,\,1$, 
\begin{equation}
    \label{relt12}
    \begin{split}
    \sqrt{\lambda_{1}\overline{\lambda_{t}}\left(\zeta\right)}+\overline{\lambda_{t1}}\left(\xi_{1}\right)>0,\hspace{0.4 cm}\sqrt{\lambda_{2}\overline{\lambda_{t}}\left(\zeta\right)}+\overline{\lambda_{t2}}\left(\xi_{2}\right)>0 
    \end{split}
\end{equation}
and
\begin{equation}
    \label{relt3}
    \begin{split}
    & 2\{\left(\lambda_{3}+\lambda_{4}c^{2}\right)\overline{\lambda_{t}}\left(\zeta\right)-\overline{\lambda_{t1}}\left(\xi_{1}\right)\overline{\lambda_{t2}}\left(\xi_{2}\right)\} \\
    & +\sqrt{\left(\lambda_{1}\overline{\lambda_{t}}\left(\zeta\right)-\overline{\lambda_{t1}}^{2}\left(\xi_{1}\right)\right)\left(\lambda_{2}\overline{\lambda_{t}}\left(\zeta\right)-\overline{\lambda_{t2}}^{2}\left(\xi_{2}\right)\right)}>0.
    \end{split}
\end{equation}
For the extremum values of $\zeta$ and $\xi_{1,2}$ we have from the inequalities in  \ref{relt12} the conditions: 
\begin{equation}
    \label{l1lt1}
    \begin{split}
        & \sqrt{\lambda_{1}\lambda_{t}}+\lambda_{t1}>0,\hspace{0.4 cm}\sqrt{\lambda_{1}\lambda_{t}}+\lambda_{t1}+\lambda_{tt1}>0,\\
        & \sqrt{\lambda_{1}\left(\lambda_{t}+\frac{\lambda_{tt}}{2}\right)}+\lambda_{t1}>0,\hspace{0.4 cm}\sqrt{\lambda_{1}\left(\lambda_{t}+\frac{\lambda_{tt}}{2}\right)}+\lambda_{t1}+\lambda_{tt1}>0;
    \end{split}
\end{equation}
and
\begin{equation}
    \label{l2lt2}
    \begin{split}
        & \sqrt{\lambda_{2}\lambda_{t}}+\lambda_{t2}>0,\hspace{0.4 cm}\sqrt{\lambda_{2}\lambda_{t}}+\lambda_{t2}+\lambda_{tt2}>0,\\
        & \sqrt{\lambda_{2}\left(\lambda_{t}+\frac{\lambda_{tt}}{2}\right)}+\lambda_{t2}>0,\hspace{0.4 cm}\sqrt{\lambda_{2}\left(\lambda_{t}+\frac{\lambda_{tt}}{2}\right)}+\lambda_{t2}+\lambda_{tt2}>0.
    \end{split}
\end{equation}
In the same way, the inequality in \ref{relt3} leads to the conditions:  
\begin{equation}
    \begin{split}
      & 2({\lambda_{3}}{\lambda_{t}}-{\lambda_{t1}}{\lambda_{t2}})  +\sqrt{\left(\lambda_{1}{\lambda_{t}}-\lambda_{t1}^{2}\right)\left(\lambda_{2}{\lambda_{t}}-\lambda_{t2}^{2}\right)}>0,  
    \end{split}
\end{equation}

\begin{equation}
    \begin{split}
      & 2({\lambda_{3}}{\lambda_{t}}-{\lambda_{t1}}{(\lambda_{t2}+\lambda_{tt2})})  +\sqrt{\left(\lambda_{1}{\lambda_{t}}-\lambda_{t1}^{2}\right)\left(\lambda_{2}{\lambda_{t}}-(\lambda_{t2}+\lambda_{tt2})^{2}\right)}>0,  
    \end{split}
\end{equation}

\begin{equation}
    \begin{split}
      & 2({\lambda_{3}}{\lambda_{t}}-(\lambda_{t1}+\lambda_{tt1}){\lambda_{t2}})  +\sqrt{\left(\lambda_{1}{\lambda_{t}}-(\lambda_{t1}+\lambda_{tt1})^{2}\right)\left(\lambda_{2}{\lambda_{t}}-\lambda_{t2}^{2}\right)}>0 , 
    \end{split}
\end{equation}

\begin{equation}
    \begin{split}
      & 2({\lambda_{3}}{\lambda_{t}}-(\lambda_{t1}+\lambda_{tt1})(\lambda_{t2}+\lambda_{tt2}))  \\
      & +\sqrt{\left(\lambda_{1}{\lambda_{t}}-(\lambda_{t1}+\lambda_{tt1})^{2}\right)\left(\lambda_{2}{\lambda_{t}}-(\lambda_{t2}+\lambda_{tt2})^{2}\right)}>0,  
    \end{split}
\end{equation}

\begin{equation}
    \begin{split}
      & 2({\lambda_{3}}(\lambda_{t}+\frac{\lambda_{tt}}{2})-{\lambda_{t1}}{\lambda_{t2}})  \\ 
      & +\sqrt{\left(\lambda_{1}(\lambda_{t}+\frac{\lambda_{tt}}{2})-\lambda_{t1}^{2}\right)\left(\lambda_{2}(\lambda_{t}+\frac{\lambda_{tt}}{2})-\lambda_{t2}^{2}\right)}>0,  
    \end{split}
\end{equation}

\begin{equation}
    \begin{split}
      & 2({\lambda_{3}}(\lambda_{t}+\frac{\lambda_{tt}}{2})-{\lambda_{t1}}(\lambda_{t2}+\lambda_{tt2}))  \\ 
      & +\sqrt{\left(\lambda_{1}(\lambda_{t}+\frac{\lambda_{tt}}{2})-\lambda_{t1}^{2}\right)\left(\lambda_{2}(\lambda_{t}+\frac{\lambda_{tt}}{2})-(\lambda_{t2}+\lambda_{tt2})^{2}\right)}>0,  
    \end{split}
\end{equation}

\begin{equation}
    \begin{split}
      & 2({\lambda_{3}}(\lambda_{t}+\frac{\lambda_{tt}}{2})-(\lambda_{t1}+\lambda_{tt1}){\lambda_{t2}})  \\ 
      & +\sqrt{\left(\lambda_{1}(\lambda_{t}+\frac{\lambda_{tt}}{2})-(\lambda_{t1}+\lambda_{tt1})^{2}\right)\left(\lambda_{2}(\lambda_{t}+\frac{\lambda_{tt}}{2})-\lambda_{t2}^{2}\right)}>0,  
    \end{split}
\end{equation}

\begin{equation}
    \begin{split}
      & 2({\lambda_{3}}(\lambda_{t}+\frac{\lambda_{tt}}{2})-(\lambda_{t1}+\lambda_{tt1})(\lambda_{t2}+\lambda_{tt2}))  \\ 
      & +\sqrt{\left(\lambda_{1}(\lambda_{t}+\frac{\lambda_{tt}}{2})-(\lambda_{t1}+\lambda_{tt1})^{2}\right)\left(\lambda_{2}(\lambda_{t}+\frac{\lambda_{tt}}{2})-(\lambda_{t2}+\lambda_{tt2})^{2}\right)}>0,  
    \end{split}
\end{equation}

\begin{equation}
    \begin{split}
      & 2((\lambda_{3}+\lambda_4){\lambda_{t}}-{\lambda_{t1}}{\lambda_{t2}})  +\sqrt{\left(\lambda_{1}{\lambda_{t}}-\lambda_{t1}^{2}\right)\left(\lambda_{2}{\lambda_{t}}-\lambda_{t2}^{2}\right)}>0,  
    \end{split}
\end{equation}

\begin{equation}
    \begin{split}
      & 2((\lambda_{3}+\lambda_4){\lambda_{t}}-{\lambda_{t1}}{(\lambda_{t2}+\lambda_{tt2})})  +\sqrt{\left(\lambda_{1}{\lambda_{t}}-\lambda_{t1}^{2}\right)\left(\lambda_{2}{\lambda_{t}}-(\lambda_{t2}+\lambda_{tt2})^{2}\right)}>0,  
    \end{split}
\end{equation}

\begin{equation}
    \begin{split}
      & 2((\lambda_{3}+\lambda_4){\lambda_{t}}-(\lambda_{t1}+\lambda_{tt1}){\lambda_{t2}})  +\sqrt{\left(\lambda_{1}{\lambda_{t}}-(\lambda_{t1}+\lambda_{tt1})^{2}\right)\left(\lambda_{2}{\lambda_{t}}-\lambda_{t2}^{2}\right)}>0 , 
    \end{split}
\end{equation}

\begin{equation}
    \begin{split}
      & 2((\lambda_{3}+\lambda_4){\lambda_{t}}-(\lambda_{t1}+\lambda_{tt1})(\lambda_{t2}+\lambda_{tt2})) \\ 
      & +\sqrt{\left(\lambda_{1}{\lambda_{t}}-(\lambda_{t1}+\lambda_{tt1})^{2}\right)\left(\lambda_{2}{\lambda_{t}}-(\lambda_{t2}+\lambda_{tt2})^{2}\right)}>0 , 
    \end{split}
\end{equation}

\begin{equation}
    \begin{split}
      & 2((\lambda_{3}+\lambda_4)(\lambda_{t}+\frac{\lambda_{tt}}{2})-{\lambda_{t1}}{\lambda_{t2}}) \\
      & +\sqrt{\left(\lambda_{1}(\lambda_{t}+\frac{\lambda_{tt}}{2})-\lambda_{t1}^{2}\right)\left(\lambda_{2}(\lambda_{t}+\frac{\lambda_{tt}}{2})-\lambda_{t2}^{2}\right)}>0,  
    \end{split}
\end{equation}

\begin{equation}
    \begin{split}
      & 2((\lambda_{3}+\lambda_4)(\lambda_{t}+\frac{\lambda_{tt}}{2})-{\lambda_{t1}}(\lambda_{t2}+\lambda_{tt2})) \\
      & +\sqrt{\left(\lambda_{1}(\lambda_{t}+\frac{\lambda_{tt}}{2})-\lambda_{t1}^{2}\right)\left(\lambda_{2}(\lambda_{t}+\frac{\lambda_{tt}}{2})-(\lambda_{t2}+\lambda_{tt2})^{2}\right)}>0,  
    \end{split}
\end{equation}

\begin{equation}
    \begin{split}
      & 2((\lambda_{3}+\lambda_4)(\lambda_{t}+\frac{\lambda_{tt}}{2})-(\lambda_{t1}+\lambda_{tt1}){\lambda_{t2}}) \\
      & +\sqrt{\left(\lambda_{1}(\lambda_{t}+\frac{\lambda_{tt}}{2})-(\lambda_{t1}+\lambda_{tt1})^{2}\right)\left(\lambda_{2}(\lambda_{t}+\frac{\lambda_{tt}}{2})-\lambda_{t2}^{2}\right)}>0,  
    \end{split}
\end{equation}

\begin{equation}
    \begin{split}
      & 2((\lambda_{3}+\lambda_4)(\lambda_{t}+\frac{\lambda_{tt}}{2})-(\lambda_{t1}+\lambda_{tt1})(\lambda_{t2}+\lambda_{tt2}))  \\ 
      & +\sqrt{\left(\lambda_{1}(\lambda_{t}+\frac{\lambda_{tt}}{2})-(\lambda_{t1}+\lambda_{tt1})^{2}\right)\left(\lambda_{2}(\lambda_{t}+\frac{\lambda_{tt}}{2})-(\lambda_{t2}+\lambda_{tt2})^{2}\right)}>0.  
    \end{split}
\end{equation}


\section{Anomaly Cancellation}
\label{sec:app0}

In this section we will describe how the gauge anomalies can be cancelled in our model which simply adds an abelian gauge group to the SM.
Taking the SM fermion charges under the new weak hypercharge, $Y^{\prime}$, to be $l$ (left-handed leptons), $q$ (left-handed quarks), $e$ (right-handed leptons), $u$ (right-handed quarks with positive isospin), $d$ (right-handed quarks with positive isospin). Let us consider each relevant gauge anomaly individually,

\begin{description}

\item[\text{$ \left[ SU(3)_c \right] ^2 U(1)_X $}]: 
\begin{equation*}
\mathcal{A} = \text{Tr} \left[ \left\{ \frac{\lambda ^a}{2}, \frac{\lambda ^b}{2} \right\} Y ' _R \right] - \text{Tr} \left[ \left\{ \frac{\lambda ^a}{2}, \frac{\lambda ^b}{2} \right\} Y ' _L \right]
\end{equation*}
\begin{equation*}
\mathcal{A} \propto \sum _{\text{quarks}} Y ' _R - \sum _{\text{quarks}} Y ' _L = \left[ 3 u + 3 d \right] - \left[ 3 \cdot 2 q \right] = 0.
\end{equation*}
Hence, 
\begin{equation}
u + d - 2 q = 0 .
\label{anomalycond1}
\end{equation}

\item[\text{$ \left[ SU(2)_L \right] ^2 U(1)_X $}]: 
\begin{equation*}
\mathcal{A} = - \text{Tr} \left[ \left\{ \frac{\sigma ^a}{2}, \frac{\sigma ^b}{2} \right\} Y ' _L \right] \propto - \sum Y_L = - \left[ 2 l + 3 \cdot 2 q \right] = 0.
\end{equation*}
Thus, 
\begin{equation}
l = - 3 q.
\label{anomalycond2}
\end{equation}

\item[\text{$ \left[ U(1)_Y \right] ^2 U(1)_X $}]: 
\begin{equation*}
\mathcal{A} = \text{Tr} \left[ \left\{ Y_R, Y_R \right\} Y ' _R \right] - \text{Tr} \left[ \left\{ Y_L, Y_L \right\} Y ' _L \right] \propto \sum  Y_R ^2 Y ' _R - \sum Y_L ^2 Y ' _L
\end{equation*}
\begin{equation*}
\mathcal{A} \propto \left[ \left( -2 \right) ^2 e + 3 \left( \frac{4}{3} \right) ^2 u + 3 \left( - \frac{2}{3} \right) ^2 d \right] - \left[ 2 \left( -1 \right) ^2 l + 3 \cdot 2 \left( \frac{1}{3} \right) ^2 q \right] = 0.
\end{equation*}
We conclude that, 
\begin{equation}
6 e + 8 u + 2 d - 3 l - q = 0.
\label{anomalycond3}
\end{equation}

\item[\text{$ U(1)_Y \left[ U(1)_X \right] ^2 $}]: 
\begin{equation*}
\mathcal{A} = \text{Tr} \left[ \left\{ Y ' _R, Y ' _R \right\} Y _R \right] - \text{Tr} \left[ \left\{ Y ' _L, Y ' _L \right\} Y _L \right] \propto \sum  Y_R {Y ' _R} ^2 - \sum Y_L {Y ' _L} ^2
\end{equation*}
\begin{equation*}
\mathcal{A} \propto \left[ \left( -2 \right) e ^2 + 3 \left( \frac{4}{3} \right) u ^2 + 3 \left( - \frac{2}{3} \right) d ^2 \right] - \left[ 2 \left( -1 \right)  l ^2 + 3 \cdot 2 \left( \frac{1}{3} \right) q ^2 \right] = 0.
\end{equation*}
That implies, 
\begin{equation}
- e ^2 + 2 u ^2 - d ^2 + l ^2 - q ^2 = 0.
\label{anomalycond4}
\end{equation}

\item[\text{$ \left[ U(1)_X \right] ^3 $}]: 
\begin{equation*}
\mathcal{A} = \text{Tr} \left[ \left\{ Y ' _R, Y ' _R \right\} Y ' _R \right] - \text{Tr} \left[ \left\{ Y ' _L, Y ' _L \right\} Y ' _L \right] \propto \sum {Y ' _R} ^3 - \sum {Y ' _L} ^3
\end{equation*}
\begin{equation*}
\mathcal{A} \propto \left[ e ^3 + 3 u ^3 + 3 d ^3 \right] - \left[ 2 l ^3 + 3 \cdot 2 q ^3 \right] = 0.
\end{equation*}
Consequently we get,
\begin{equation}
e ^3 + 3 u ^3 + 3 d ^3 - 2 l ^3 - 6 q ^3 = 0.
\label{anomalycond5}
\end{equation}

\end{description} 

These relations between the charges under U(1) are general, since we have not added any condition concerning the Lagrangians of the model. If we want  to accomodate neutrino masses either via type I or type II seesaw mechanisms and explain the absence of flavor changing interactions, further relations between hypercharges under the new gauge symmetry U(1) arise as summarized in eq. (\ref{expres_cargas_u_d}) and eq.(\ref{expres_cargas_d}).

\section{Scalar masses and mixings}

In this appendix we show in details how to obtain approximate expressions for the masses and mixing angles for the physical scalars in the relevant regime adopted in this paper, $v _i \sim 100 \text{\ GeV} $, $v _t \ll v _i$. In the first section we treat the CP-even scalars, and in the next we treat the charged scalars.

\subsection{CP-even scalars}
\label{sec:app1}

In general, it is not possible to obtain analytic expressions for the diagonalization of the mass matrix $M ^2 _{\text{CPeven}}$, given in eq. (\ref{cp_even_mass_matrix}). The masses of the neutral scalars (eigenvalues of that matrix) are given implicitly as the solutions of the polynomial equation,
\begin{equation}
a x ^3 + b x ^2 + c x + d = 0 ,
\end{equation}
with $a$, $b$, $c$ and $d$ given by,
\begin{equation}
a = 8 v _1 v _2 v _t ,
\end{equation}
\begin{equation}
\begin{split}
b = - 16 v _1 v _2 v _t [ \lambda _1 v _1 ^2 + \lambda _2 v _2 ^2 + v _t ^2 ( \lambda _t + \lambda _{tt} ) ] - 4 \sqrt{2} \mu _{t2} \left( v _1 ^2 v _2 ^2 + v _1 ^2 v _t ^2 + v _2 ^2 v _t ^2 \right) \\
\end{split}
\end{equation}
\begin{equation}
\begin{split}
c & = 8 v _1 v _2 v _t [ 4 \lambda _1 \lambda _2  v _1 ^2 v _2 ^2 - ( \lambda _3 + \lambda _4 ) ^2  v _1 ^2 v _2 ^2 + \left( 4 \lambda _1 \lambda _t - \lambda _{t1} ^2 \right) v _1 ^2 v _t ^2 + \left( 4 \lambda _2 \lambda _t - \lambda _{t2} ^2 \right) v _2 ^2 v _t ^2 \\ 
& + 4 \lambda _{tt} v _t^2 \left( \lambda _1 v _1 ^2 + \lambda _2 v _2 ^2 \right) - 2 v _t ^2 \left( \lambda _{t1} \lambda _{tt1} v _1 ^2 + \lambda _{t2} \lambda _{tt2} v _2 ^2 \right) - v _t ^2 \left( \lambda _{tt1} ^2 v _1 ^2 + \lambda _{tt2} ^2 v _2 ^2 \right) ] \\ 
& + 8 \sqrt{2} \mu _{t2} [ \lambda _1 v _1 ^4 \left( v _2 ^2 + v _t ^2 \right) + \lambda _2 v _2 ^4 \left( v _1 ^2 + v _t ^2 \right) + \lambda _t v _t ^4 \left( v _1 ^2 + v _2 ^2 \right) + \lambda _{tt} v _t ^4 \left( v _1 ^2 + v _2 ^2 \right) \\
& + ( \lambda _3 + \lambda _4 ) v _1 ^2 v _2 ^2 v _t ^2 + ( \lambda _{t1} + \lambda _{t2} ) v _1 ^2 v _2 ^2 v _t ^2 + ( \lambda _{tt1} + \lambda _{tt2} ) v _1 ^2 v _2 ^2 v _t ^2 ]
\end{split}
\end{equation}
\begin{equation}
\begin{split}
d & = 16 v _1^3 v _2^3 v _t^3 \{ \lambda _1 ( \lambda _{t2} + \lambda _{tt2} ) ^2 + \lambda _2 ( \lambda _{t1} + \lambda _{tt1} ) ^2 + ( \lambda _t + \lambda _{tt} ) [ ( \lambda _3 + \lambda _4 ) ^2 - 4 \lambda _1 \lambda _2 ] \\
& - ( \lambda _3 + \lambda _4 ) ( \lambda _{t1} + \lambda _{tt1} ) ( \lambda _{t2} + \lambda _{tt2} ) \} - 4 \sqrt{2} \mu _{t2} \{ [ 4 \lambda _1 \lambda _2 - ( \lambda _3 + \lambda _4 ) ^2 ] v _1 ^4 v _2 ^4 \\
& + [ 4 \lambda _2 ( \lambda _t + \lambda _{tt} ) - ( \lambda _{t2} + \lambda _{tt2} ) ^2 ] v _2 ^4 v _t ^4 + [ 4 \lambda _1 ( \lambda _t + \lambda _{tt} ) - ( \lambda _{t1} + \lambda _{tt1} ) ^2 ] v _1 ^4 v _t ^4 \} \\
& + 8 \sqrt{2} \mu _{t2} v _1 ^2 v _2 ^2 v _t ^2 \{ [ ( \lambda _{t1} + \lambda _{tt1} ) ( \lambda _{t2} + \lambda _{tt2} ) - 2 ( \lambda _3 + \lambda _4 ) ( \lambda _t + \lambda _{tt} ) ] v _t ^2 \\
& + [ ( \lambda _3 + \lambda _4 ) ( \lambda _{t1} + \lambda _{tt1} ) - 2 \lambda _1 ( \lambda _{t2} + \lambda _{tt2} ) ] v _1 ^2 + [ ( \lambda _3 + \lambda _4 ) ( \lambda _{t2} + \lambda _{tt2} ) - 2 \lambda _2 ( \lambda _{t1} + \lambda _{tt1} ) ] v _2 ^2 \} \\
& - 16 \mu _{t2} ^2 v _1 v _2 v _t [ ( \lambda _3 + \lambda _4 ) v _1 ^2 v _2 ^2 + ( \lambda _{t1} + \lambda _{tt1} ) v _1 ^2 v _t ^2 + ( \lambda _{t2} + \lambda _{tt2} ) v _2 ^2 v _t ^2 ] + 8 \sqrt{2} \mu _{t2} ^3 v _1 ^2 v _2 ^2 v _t ^2 ,
\end{split}
\end{equation}
This equation can be solved numerically once the set of parameters is fixed. For the mixing angles, it is very difficult even to furnish an equation that determine them in terms of the parameters of the potential, because of the difficulty in computing the eigenvectors of $M ^2 _{\text{CPeven}}$. 

There are some limits, however, in which these expressions are calculable. The idea is to take advantage of the different energy scales involved and decompose the original matrix into matrices whose entries belong to the same scale. 

Let us decompose $M ^2 _{\text{CPeven}}$,
\begin{equation*}
M ^2 _{\text{CPeven}} = \begin{pmatrix}  2 \lambda _1 v _1 ^2 & ( \lambda _3 + \lambda _4 ) v _1 v _2 & ( \lambda _{t1} + \lambda _{tt1} ) v _1 v _t \\
 ( \lambda _3 + \lambda _4 ) v _1 v _2 & 2 \lambda _2 v _2 ^2 & ( \lambda _{t2} + \lambda _{tt2} ) v _2 v _t - \sqrt{2} \mu _{t2} v _2 \\
 ( \lambda _{t1} + \lambda _{tt1} ) v _1 v _t & ( \lambda _{t2} + \lambda _{tt2} ) v _2 v _t - \sqrt{2} \mu _{t2} v _2 & 2 ( \lambda _t + \lambda _{tt} ) v _t ^2 + \frac{\mu _{t2} v _2^2}{\sqrt{2} v _t} 
\end{pmatrix} ,
\end{equation*}
in the following way,
\begin{equation}
\begin{split}
\label{decomposition_neutral_scalars_mass}
M ^2 _{\text{CPeven}} & = M _1 ^2 + M _2 ^2 \\
& = v _1 v _2 \begin{pmatrix}  2 \lambda _1 \frac{v _1}{v _2} & \lambda _3 + \lambda _4 & 0 \\
 \lambda _3 + \lambda _4 & 2 \lambda _2 \frac{v _2}{v _1} & 0 \\
 0 & 0 & 0 \end{pmatrix} \\
 & + \sqrt{2} \mu _{t2} v _2 \begin{pmatrix}  0 & 0 & ( \lambda _{t1} + \lambda _{tt1} ) \frac{v _1 v _t}{\sqrt{2} \mu _{t2} v _2} \\ 
 0 & 0 & ( \lambda _{t2} + \lambda _{tt2} ) \frac{v _t}{\sqrt{2} \mu _{t2}} - 1 \\
 ( \lambda _{t1} + \lambda _{tt1} ) \frac{v _1 v _t}{\sqrt{2} \mu _{t2} v _2} & ( \lambda _{t2} + \lambda _{tt2} ) \frac{v _t}{\sqrt{2} \mu _{t2}} - 1 & ( \lambda _t + \lambda _{tt} ) \frac{2 v _t ^2}{\sqrt{2} \mu _{t2} v _2} + \frac{v _2}{2 v _t} 
\end{pmatrix} .
\end{split}
\end{equation}
The matrix $ M _1 ^2 $ would be the mixing matrix of the neutral scalars in case there were just the two doublets, while $ M _2 ^2 $ account for the effects of the presence of the triplet. 
Let us first consider the limit $ v _t , \mu _{t2} \ll v _i $.   
In this case, the decomposition (\ref{decomposition_neutral_scalars_mass}) makes it clear that the triplet decouples from the doublets. The matrix $ M _1 ^2 $ remains the same and $ M _2 ^2 $ reduces to $ M _2 ^2 = \text{diag} ( 0 , 0, \mu _{t2} v _2 ^2 / \sqrt{2} v _t ) $, from which we obtain 
immediately the mass of $H _t$. To diagonalize $ M _1 ^2 $ we need only one mixing angle, so that we can make $ \alpha _1 , \alpha _2 \rightarrow 0 $ in eq.(\ref{rot_to_phy_basis_cp_even}), leading to the following physical fields,
\begin{equation*}
\begin{split}
\begin{pmatrix} h \\ H \\ H _t \end{pmatrix} = & \begin{pmatrix} c _\alpha & s _\alpha & 0 \\ - s _\alpha & c _\alpha & 0 \\ 0 & 0 & 1 \end{pmatrix} \begin{pmatrix} \rho _1 \\ \rho _2 \\ \rho _t \end{pmatrix}
\end{split}
\end{equation*}
with $\alpha$ given by,
\begin{equation}
\label{tan_2_alpha}
\tan 2 \alpha = \frac{( \lambda _3 + \lambda _4 ) v _1 v _2}{\lambda _1 v _1 ^2 - \lambda _2 v _2 ^2} .
\end{equation}
From the eigenvalues of $M _1 ^2$ and $M _2 ^2$, we have the masses,
\begin{equation}
\label{masses_neutral_app1}
m _{h,H} ^{' 2} = \lambda _1 v _1 ^2 + \lambda _2 v _2 ^2 \pm \sqrt{ ( \lambda _1 v _1 ^2 - \lambda _2 v _2 ^2 )^2 + ( \lambda _3 + \lambda _4 ) ^2 v _1 ^2 v _2 ^2 } 
\end{equation}
\begin{equation}
\label{masses_neutral_app2}
m _{H _t} ^2 = \frac{\mu _{t2} v _2^2}{\sqrt{2} v _t} ,
\end{equation}
with $ m ' _h < m ' _H $ \footnote{The prime in $m _{h,H} ^{' 2}$ was inserted here to differentiate these mass expressions from the masses $ m _{h,H} ^2 $, given in eqs. (\ref{masses_neutral_app2_1}) and (\ref{masses_neutral_app2_2}), which are the masses of $h$ and $H$ calculated in another limit.}. \\

Now, relaxing the condition on $ \mu _{t2} $ and allowing it to increase to the same order of $ v _i $ or higher, this comparatively large value of $ \mu _{t2} $ produces a sizable perturbation on the spectrum obtained above, but as we will see, the mass expressions are somewhat similar to the ones obtained in eqs. (\ref{masses_neutral_app1}) and (\ref{masses_neutral_app2}). In this case, is still possible to diagonalize the matrices $ M _1 ^2 $ and $ M _2 ^2 $ almost independently.
First, as $v _t$ is always taken to be small but now $ \mu _{t2} $ can be large, $ M _2 ^2 $ can be approximated by,
\begin{equation}
\begin{split}
M _2 ^2 & = \sqrt{2} \mu _{t2} v _2 \begin{pmatrix}  0 & 0 & 0 \\ 
 0 & 0 & - 1 \\
 0 & - 1 & \frac{v _2}{2 v _t} 
\end{pmatrix} .
\end{split}
\end{equation}
$ M _2 ^2 $ is diagonalized by moving to an intermediate basis $ ( H _1 , H _2 , H _3 ) $ through a rotation $ R _{\alpha _2} $,
\begin{equation}
\begin{pmatrix} H _1 \\ H _2 \\ H _3 \end{pmatrix} = \begin{pmatrix} 1 & 0 & 0 \\ 0 & c _{\alpha _2} & s _{\alpha _2}  \\ 0 & - s _{\alpha _2} & c _{\alpha _2} \end{pmatrix}  \begin{pmatrix} \rho _1 \\ \rho _2 \\ \rho _t \end{pmatrix} ,
\end{equation}
with,
\begin{equation}
\sin \alpha _2 = \frac{2 v _t}{v _2} \text{\ \ \ , \ \ \ \ } \cos \alpha _2 \simeq 1 ,
\end{equation}
so that,
\begin{equation}
\begin{split}
M _{2\text{diag}} ^2 & = R _{\alpha _2} M _2 ^2 R _{\alpha _2} ^T \\
& = \sqrt{2} \mu _{t2} v _2 \begin{pmatrix}  0 & 0 & 0 \\  0 & - \frac{2 v _t}{v _2} & O ( \frac{v _t ^2}{v _2 ^2} ) \\  0 & O ( \frac{v _t ^2}{v _2 ^2} ) & \frac{v _2}{2 v _t} \end{pmatrix} .
\end{split}
\end{equation}

The effect of this rotation on $ M _1 ^2 $ is
\begin{equation}
\begin{split}
R _{\alpha _2} M _1 ^2 R _{\alpha _2} ^T & = v _1 v _2 \begin{pmatrix} 2 \lambda _1 \frac{v _1}{v _2} & ( \lambda _3 + \lambda _4 ) c _{\alpha _2} & - ( \lambda _3 + \lambda _4 ) s _{\alpha _2} \\
 ( \lambda _3 + \lambda _4 ) c _{\alpha _2} & 2 \lambda _2 \frac{v _2}{v _1} c _{\alpha _2} ^2 & - 2 \lambda _2 \frac{v _2}{v _1} s _{\alpha _2} c _{\alpha _2} \\
 - ( \lambda _3 + \lambda _4 ) s _{\alpha _2} & - 2 \lambda _2 \frac{v _2}{v _1} s _{\alpha _2} c _{\alpha _2} & 2 \lambda _2 \frac{v _2}{v _1} s _{\alpha _2} ^2 \end{pmatrix} .
\end{split}
\end{equation}
As $ \sin \alpha _2 \ll 1 $, at leading order we have $ R _{\alpha _2} M _1 ^2 R _{\alpha _2} ^T \simeq M _1 ^2 $, and the rotation $ R _{\alpha _2} $ does not change $ M _1 ^2 $.  
Then, rotating $ M _1 ^2 $ by $ R _\alpha $, with $ \alpha $ given in eq. (\ref{tan_2_alpha}), we move to the physical basis,
\begin{equation}
\begin{pmatrix} h \\ H \\ H _t \end{pmatrix} = \begin{pmatrix} c _\alpha & s _\alpha & 0 \\ - s _\alpha & c _\alpha & 0 \\ 0 & 0 & 1 \end{pmatrix}  \begin{pmatrix} H _1 \\ H _2 \\ H _3 \end{pmatrix} ,
\end{equation}
such that,
\begin{equation}
\begin{split}
M _{1\text{diag}} ^2 & = R _\alpha M _2 ^2 R _\alpha ^T \\
& = \begin{pmatrix} m _h ^{'2} & 0 & 0 \\ 0 & m _H ^{' 2} & 0 \\ 0 & 0 & 0 \end{pmatrix} ,
\end{split}
\end{equation}
where $ m _h ^{'2} $ and $ m _H ^{'2} $ are as given in eq. (\ref{masses_neutral_app1}). This second rotation $ R _\alpha $, albeit diagonalize $ M _1 ^2 $, tends to disturb the previous diagonalization of $ M _2 ^2 $, by generating corrections to the diagonal elements and also off-diagonal elements in $ M _{2\text{diag}} ^2 $,
\begin{equation}
M _{2\text{diag}} ^{' 2} = R _\alpha M _{2\text{diag}} ^2 R _\alpha ^T = \sqrt{2} \mu _{t2} v _2 \begin{pmatrix} - 2 {s _{\alpha}} ^2 \frac{v _t}{v _2} & O ( \frac{v _t}{v _2} ) & 0 \\  O ( \frac{v _t}{v _2} ) & - 2 {c _{\alpha}} ^2 \frac{v _t}{v _2} & 0 \\  0 & 0 & \frac{v _2}{2 v _t} \end{pmatrix} .
\end{equation}
This matrix is diagonal up to $O (v _t / v _2)$ terms, which can be discarded for sufficiently small $v _t$. As we are interested in extracting the leading order correction to the masses of $h$ and $H$, we shall keep the diagonal elements.

In summary what we have done is,
\begin{equation*}
\begin{split}
\boldsymbol{\rho} ^T M _{\text{CPeven}} ^2 \boldsymbol{\rho} & = \boldsymbol{\rho} ^T ( M _1 ^2 + M _2 ^2 ) \boldsymbol{\rho} \\
& = \boldsymbol{H} _1 ^T ( R _{\alpha _2} M _1 ^2 R _{\alpha _2} ^T + R _{\alpha _2} M _2 ^2 R _{\alpha _2} ^T ) \boldsymbol{H} _1 \\
& \simeq \boldsymbol{H} _1 ^T ( M _1 ^2 + M _{2\text{diag}} ^2 ) \boldsymbol{H} _1 \\
& = \boldsymbol{H} ^T ( R _\alpha M _1 ^2 R _\alpha ^T + R _\alpha M _{2diag} ^2 R _\alpha ^T ) \boldsymbol{H} \\
\boldsymbol{\rho} ^T M _{\text{CPeven}} ^2 \boldsymbol{\rho} & \simeq \boldsymbol{H} ^T ( M _{1diag} ^2 + M _{2diag} ^{' 2} ) \boldsymbol{H} .
\end{split}
\end{equation*}
From,
\begin{equation}
\begin{split}
M _{1diag} ^2 + M _{2diag} ^{' 2} & = \begin{pmatrix} m _h ^{'2} - 2 \sqrt{2} {s _{\alpha}} ^2 \mu _{t2} v _t & 0 & 0 \\  0 & m_H ^{'2} - 2 \sqrt{2} {c _{\alpha}} ^2 \mu _{t2} v _t & 0 
\\  0 & 0 
& \frac{\mu _{t2} v _2 ^2}{\sqrt{2} v _t} \end{pmatrix} ,
\end{split}
\end{equation}
we can read the scalar masses,
\begin{equation}
\label{masses_neutral_app2_1}
m _{h} ^2 = \lambda _1 v _1 ^2 + \lambda _2 v _2 ^2 - \sqrt{ ( \lambda _1 v _1 ^2 - \lambda _2 v _2 ^2 )^2 + ( \lambda _3 + \lambda _4 ) ^2 v _1 ^2 v _2 ^2 } - 2 \sqrt{2} \sin ^2 \alpha \text{\ } \mu _{t2} v _t
\end{equation}
\begin{equation}
\label{masses_neutral_app2_2}
m _{H} ^2 = \lambda _1 v _1 ^2 + \lambda _2 v _2 ^2 + \sqrt{ ( \lambda _1 v _1 ^2 - \lambda _2 v _2 ^2 )^2 + ( \lambda _3 + \lambda _4 ) ^2 v _1 ^2 v _2 ^2 } - 2 \sqrt{2} \cos ^2 \alpha \text{\ } \mu _{t2} v _t
\end{equation}
\begin{equation}
\label{masses_neutral_app2_3}
m _{H _t} ^2 = \frac{\mu _{t2} v _2 ^2}{\sqrt{2} v _t} .
\end{equation} 
Note that the expressions obtained agree with those given in eq. (\ref{masses_neutral_app1})-(\ref{masses_neutral_app2}) if we take $ \mu _{t2} \ll v _i $, as expected. The main difference in the expressions in these two limits is the presence of the correction terms $ - 2 \sqrt{2} \sin ^2 \alpha \mu _{t2} v _t $ and $ - 2 \sqrt{2} \cos ^2 \alpha \mu _{t2} v _t $ in $ m _{h} ^2 $ and $ m _{H} ^2 $, respectively, which pushes down their values, making $ h $ and $ H $ lighter than it would be in the absence of the triplet. 
Notice also that in this approximation, we managed to perform the diagonalization using only two mixing angles, $ \alpha $ and $ \alpha _2 $, instead of the three angles needed in the general case.

\subsection{Charged scalars}
\label{sec:app2}

In this section we will apply to the charged scalars mass matrix the same method used in the previous section for the neutral scalars. First note that the matrix (\ref{charged_scalars_mass_matrix}),
\begin{equation*}
M _{\text{Charged}} ^2 = \frac{1}{2} \begin{pmatrix} - \lambda _4 v _2 ^2 - \lambda _{tt1} v _t ^2 & \lambda _4 v _1 v _2 & \lambda _{tt1} v _1 v _t / \sqrt{2} \\
 \lambda _4 v _1 v _2 & - \lambda _4 v _1 ^2 - \lambda _{tt2} v _t ^2 + 2 \sqrt{2} \mu _{t2} v _t & \frac{1}{2} ( \sqrt{2} \lambda _{tt2} v _t - 4 \mu _{t2} ) v _2 \\
 v _1 v _t \lambda _{tt1} / \sqrt{2} & \frac{1}{2} ( \sqrt{2} \lambda _{tt2} v _t - 4 \mu _{t2} ) v _2 & \frac{ \sqrt{2} \mu _{t2} v _2 ^2 }{v _t} - \frac{1}{2} ( \lambda _{tt1} v _1 ^2 + \lambda _{tt2} v _2 ^2 )
\end{pmatrix} ,
\end{equation*}
can be decomposed as, 
\begin{equation}
\begin{split}
M _{\text{Charged}} ^2 & = M _{1+} ^2 + M _{2+} ^2 \\
& = \frac{1}{2} \begin{pmatrix} - \lambda _4 v _2 ^2 & \lambda _4 v _1 v _2 & 0 \\
 \lambda _4 v _1 v _2 & - \lambda _4 v _1 ^2 & 0 \\
 0 & 0 & 0 \end{pmatrix} \\
 & + \frac{1}{2} \begin{pmatrix} - \lambda _{tt1} v _t ^2 & 0 & \lambda _{tt1} v _1 v _t / \sqrt{2} \\
 0 & - \lambda _{tt2} v _t ^2 + 2 \sqrt{2} \mu _{t2} v _t & \frac{1}{2} ( \sqrt{2} \lambda _{tt2} v _t - 4 \mu _{t2} ) v _2 \\
 v _1 v _t \lambda _{tt1} / \sqrt{2} & \frac{1}{2} ( \sqrt{2} \lambda _{tt2} v _t - 4 \mu _{t2} ) v _2 & \frac{ \sqrt{2} \mu _{t2} v _2 ^2 }{v _t} - \frac{1}{2} ( \lambda _{tt1} v _1 ^2 + \lambda _{tt2} v _2 ^2 )
\end{pmatrix} .
\end{split}
\end{equation}
In the matrix $ M_{2+}^2 $ are contained all the mixing effects among the doublets and the triplet. If $ M_{2+}^2 $ vanished, there would be mixing only between the two doublets, as described by $ M_{1+}^2 $, leading to a charged Goldstone boson and a charged physical scalar, as in the usual 2HDM. Let's again consider the limit $ v _t , \mu _{t2} \ll v _i $, in which $ M _{1+} ^2 $ remains the same and $ M _{2+} ^2 $ reduces to $ M _{2+}^2 = \text{diag} ( 0 , 0, \sqrt{2} \mu _{t2} v _2 ^2 / 2 v _t - \lambda _{tt1} v _1 ^2 / 2 - \lambda _{tt2} v _2 ^2 / 2 ) $, so that the triplet completely decouples from the doublets, which still mix with themselves. In this case, it is necessary only one angle in the diagonalization and we can make $ \beta _1 , \beta _2 \rightarrow 0 $ in eq.(\ref{rot_to_phy_basis_charged_scalars}), leading to the following physical fields,
\begin{equation*}
\begin{split}
\begin{pmatrix} G ^+ \\ H ^+ \\ H _t ^+ \end{pmatrix} = & \begin{pmatrix} c _\beta & s _\beta & 0 \\ - s _\beta & c _\beta & 0 \\ 0 & 0 & 1 \end{pmatrix} \begin{pmatrix} \phi _1 ^+ \\ \phi _2 ^+ \\ \Delta ^+ \end{pmatrix}
\end{split}
\end{equation*}
with,
\begin{equation}
\label{tan_2_beta}
\tan 2 \beta = \frac{2 v _1 v _2}{v _1 ^2 - v _2 ^2}  ,
\end{equation}
which can be put in the form $ \tan 2 \beta = 2 \tan \beta / ( 1 - \tan ^2 \beta ) $ by dividing numerator and denominator by $ v _1 ^2 $, so that
\begin{equation}
\label{tan_beta}
\tan \beta = \frac{v_2}{v _1} .
\end{equation}
The masses of $ H ^+ $ and $ H _t ^+ $ in this approximation are,
\begin{equation}
m _{H ^+} ^2 = - \frac{1}{2} \lambda _4 v ^2 ,
\end{equation}
\begin{equation}
m _{H _t ^+} ^2 = \frac{\mu _{t2} v _2 ^2 }{\sqrt{2} v _t} - \frac{1}{4} ( \lambda _{tt1} v _1 ^2 + \lambda _{tt2} v _2 ^2 ) .
\end{equation}
As mentioned early, for $ \mu _{t2} \simeq v _t $ the mass of $ H _t ^+ $ is small and may be in tension with existing bounds, so that in this limit $ \mu _{t2} / v _t > 1 $ is favored.

Now, allowing $\mu _{t2}$ be large but keeping $v _t$ small, $ M _{2+} ^2 $ reduces to,
\begin{equation}
\begin{split}
M _{2+} ^2 & = \mu _{t2} \begin{pmatrix} 0 & 0 & 0 \\
 0 & \sqrt{2} v _t & - v _2 \\
 0 & - v _2 & \frac{ v _2 ^2 }{\sqrt{2} v _t} 
\end{pmatrix} .
\end{split}
\end{equation}
$ M _{2+} ^2 $ is diagonalized by moving to an intermediate basis $ ( H _1 ^+ , H _2 ^+ , H _3 ^+ ) $ through a rotation $ R _{\beta _2} $,
\begin{equation}
\begin{pmatrix} H _1 ^+ \\ H _2 ^+ \\ H _3 ^+ \end{pmatrix} = \begin{pmatrix} 1 & 0 & 0 \\ 0 & c _{\beta _2} & s _{\beta _2}  \\ 0 & - s _{\beta _2} & c _{\beta _2} \end{pmatrix} \begin{pmatrix} \phi _1 ^+ \\ \phi _2 ^+ \\ \Delta ^+ \end{pmatrix} ,
\end{equation}
so that,
\begin{equation}
R _{\beta _2} M _{2+} ^2 R _{\beta _2} ^T = M _{2+\text{diag}} ^2 ,
\end{equation}
with,
\begin{equation}
\sin \beta _2 = \frac{\sqrt{2} v _t}{\sqrt{v _2 ^2 + 2 v _t ^2}} \simeq \frac{\sqrt{2} v _t}{v _2} ,
\end{equation}
and,
\begin{equation}
\cos \beta _2 = \frac{v _2}{\sqrt{v _2 ^2 + 2 v _t ^2}} \simeq 1 .
\end{equation}

The effect of this rotation on $ M _{1+} ^2 $ is
\begin{equation}
\begin{split}
R _{\beta _2} M _{1+} ^2 R _{\beta _2} ^T & = \frac{1}{2} \begin{pmatrix} - \lambda _4 v _2 ^2 & \lambda _4 v _1 v _2 c _{\beta _2} & \lambda _4 v _1 v _2 s _{\beta _2} \\
 \lambda _4 v _1 v _2 c _{\beta _2} & - \lambda _4 v _1 ^2 c _{\beta _2} ^2 + \frac{ \lambda _{tt1} v _1 ^2 + \lambda _{tt2} v _2 ^2 }{2} s _{\beta _2} ^2 & - ( \lambda _4 v _1 ^2 + \frac{ \lambda _{tt1} v _1 ^2 + \lambda _{tt2} v _2 ^2 }{2} ) s _{\beta _2} c _{\beta _2} \\
 - \lambda _4 v _1 v _2 s _{\beta _2} & ( \lambda _4 v _1 ^2 + \frac{ \lambda _{tt1} v _1 ^2 + \lambda _{tt2} v _2 ^2 }{2} ) s _{\beta _2} c _{\beta _2} & - \frac{ \lambda _{tt1} v _1 ^2 + \lambda _{tt2} v _2 ^2 }{2} s _{\beta _2} ^2 + \lambda _4 v _1 ^2 c _{\beta _2} ^2 \end{pmatrix}  
\end{split}
\end{equation}
As $ \sin \beta _2 \ll 1 $, at leading order we have $ R _{\beta _2} M _{1+} ^2 R _{\beta _2} ^T \simeq M _{1+} ^2 $, and the rotation $ R _{\beta _2} $ does not change $ M _{1+} ^2 $, as we wanted. Then, rotating $ M _{1+} ^2 $ by a matrix $ R _\beta $, with $ \beta $ given by eq. (\ref{tan_beta}), we move to the physical basis,
\begin{equation}
\begin{pmatrix} G _1 ^+ \\ H ^+ \\ H _t ^+ \end{pmatrix} = \begin{pmatrix} c _\beta & s _\beta & 0 \\ - s _\beta & c _\beta & 0 \\ 0 & 0 & 1 \end{pmatrix} \begin{pmatrix} H _1 ^+ \\ H _2 ^+ \\ H _3 ^+ \end{pmatrix} .
\end{equation}
Note that this second rotation $ R _\beta $ does not disturb the diagonalization of $ M _{2+} ^2 $,
\begin{equation}
R _\beta M _{2+\text{diag}} ^2 R _\beta ^T = \begin{pmatrix} c _\beta & s _\beta & 0 \\ - s _\beta & c _\beta & 0 \\ 0 & 0 & 1 \end{pmatrix} \begin{pmatrix} 0 & 0 & 0 \\ 0 & 0 & 0 \\ 0 & 0 & \frac{\mu _{t2} v _2 ^2}{\sqrt{2} v _t} \end{pmatrix} \begin{pmatrix} c _\beta & - s _\beta & 0 \\ s _\beta & c _\beta & 0 \\ 0 & 0 & 1 \end{pmatrix} = \begin{pmatrix} 0 & 0 & 0 \\ 0 & 0 & 0 \\ 0 & 0 & \frac{\mu _{t2} v _2 ^2}{\sqrt{2} v _t} \end{pmatrix} = M _{2+\text{diag}} ^2 .
\end{equation}

Thus, the diagonalization of $M _{1+} ^2$ and $M _{2+} ^2$ leads to,
\begin{equation}
\begin{split}
M _{1+diag} ^2 + M _{2+diag} ^2 & = \begin{pmatrix} 0 & 0 & 0 \\  0 & - \frac{1}{2} \lambda _4 v ^2 & 0 \\  0 & 0 & \frac{\mu _{t2} v _2 ^2}{\sqrt{2} v _t} \end{pmatrix} ,
\end{split}
\end{equation}
from which we obtain the masses for $H ^+$ and $H _t ^+$,
\begin{equation}
\label{masses_charged_app2_1}
m _{H ^+} ^2 = - \frac{1}{2} \lambda _4 v ^2 ,
\end{equation}
\begin{equation}
\label{masses_charged_app2_2}
m _{H _t ^+} ^2 = \frac{\mu _{t2} v _2 ^2}{\sqrt{2} v _t} .
\end{equation}
As a consistency check, notice that taking the limit $v _t \ll v _i$ directly in the eqs. (\ref{masses_charged_exact_1}) and (\ref{masses_charged_exact_2}), we obtain as result the eqs. (\ref{masses_charged_app2_1}) and (\ref{masses_charged_app2_2}). Finally, note that for the diagonalization in this limit we need to use only two mixing angles, $ \beta $ and $ \beta _2 $, instead of the three angles needed in the general case.

\bibliographystyle{JHEP}
\bibliography{referencias}

\end{document}